\documentclass[prd,aps,twocolumn,a4paper,showkeys,nofootinbib,floatfix]{revtex4-1}

\usepackage{graphicx,psfrag}
\usepackage{epstopdf}
\usepackage{mathrsfs}
\usepackage{amsmath,amsfonts,amssymb}
\usepackage{multirow}
\usepackage{comment}
\usepackage{tipa}
\usepackage[normalem]{ulem}
\usepackage{color}
\usepackage{hyperref}
\usepackage{enumitem}
\usepackage{tipa}
\usepackage{bm}
\usepackage{algorithm}
\usepackage{algpseudocode}
\usepackage{lipsum}
\usepackage{aas_macros}
\usepackage{array}
\newcommand{\be}{\begin{equation}}
\newcommand{\ee}{\end{equation}}
\newcommand{\bea}{\begin{eqnarray}}
\newcommand{\eea}{\end{eqnarray}}
\newcommand{\bel}{\begin{align}}
\newcommand{\eel}{\end{align}}

\hypersetup{
    colorlinks=true,
    linkcolor=red,
    citecolor=blue,
    urlcolor=black
}

\def\rm{\mathrm}
\def\m0{m_\rm{0}}
\def\mgamma{m_\rm{\gamma}}
\def\logmgamma{\log_{10}\mgamma}
\def\Rd{R_\rm{d}}
\def\zd{z_\rm{d}}
\def\rb{r_\rm{b}}
\def\fb{f_\rm{b}}
\def\N{N}
\def\logN{\log_{10}\N}
\def\A{\mathcal{A}}
\def\O{\mathcal{O}}
\def\SBIres{\rm{SBI}_\rm{Res}}
\def\SBIunres{\rm{SBI}_\rm{Unres}}

\def\Msun{\mathrm{M}_{\odot}}

\usepackage{pifont}

\usepackage{color}
\definecolor{cyan}{rgb}{0,0.6,0.6}
\definecolor{orange}{rgb}{0.9,0.5,0}
\definecolor{magenta}{rgb}{1,0,1}
\definecolor{purple}{rgb}{0.8,0.4,0.8}
\definecolor{gray}{rgb}{0.8242,0.8242,0.8242}

\definecolor{nicegreen}{rgb}{0.1,0.5,0.1}
\definecolor{nicered}{rgb}{0.7,0.1,0.1}

\begin{document}

\title{Simulation-based population inference of LISA's Galactic binaries: \\Bypassing the global fit}

 \author{Rahul \surname{Srinivasan}${}^{1,2}$}
 \author{Enrico \surname{Barausse}${}^{1,2}$}
 \author{Natalia \surname{Korsakova}${}^{3,4}$}
 \author{Roberto \surname{Trotta}${}^{1,2,5,6}$}

\affiliation{${}^{1}$SISSA, Via Bonomea 265, 34136 Trieste, Italy and INFN Sezione di Trieste}
\affiliation{${}^{2}$IFPU - Institute for Fundamental Physics of the Universe, Via Beirut 2, 34014 Trieste, Italy}
 \affiliation{${}^{3}$Universit\'e C\^ote d’Azur, Observatoire de la C\^ote d’Azur, CNRS, Artemis, Bd de l'Observatoire, F-06304 Nice, France}
\affiliation{${}^{4}$
 Astroparticule et Cosmologie, CNRS, Universit\'e Paris Cit\'e, F-75013 Paris, France
}%
\affiliation{${}^{5}$Department of Physics, Imperial College London, SW7 2AZ London, UK}
\affiliation{${}^{6}$Italian Research Center on High Performance Computing, Big Data and Quantum Computing}

\date{\today}

\begin{abstract}
The Laser Interferometer Space Antenna (LISA) is expected to detect thousands of individually resolved gravitational wave sources, overlapping in time and frequency, on top of unresolved astrophysical and/or primordial backgrounds. Disentangling resolved sources from backgrounds and extracting their parameters in a computationally intensive ``global fit'' is normally regarded as a necessary step toward reconstructing the properties of the underlying astrophysical populations. Here, we show that it is in principle feasible to infer the population properties of the most numerous of LISA sources -- Galactic double white dwarfs -- directly from the frequency (or, equivalently, time) strain series by adopting a simulation-based approach, without extracting and estimating the parameters of each single source. By training a normalizing flow on a custom-designed compression of simulated LISA frequency series from the Galactic double white dwarf population, we demonstrate how to infer the posterior distribution of population parameters (e.g., mass function, frequency, and spatial distributions). This allows for extracting information on the population parameters from both resolved and unresolved sources simultaneously and in a computationally efficient manner. This approach can be extended to other source classes (e.g., massive and stellar-mass black holes, extreme mass ratio inspirals) and to scenarios involving non-Gaussian or non-stationary noise (e.g., data gaps), provided that fast and accurate simulations are available. 
\end{abstract}

\maketitle

\section{Introduction} 
\label{sec:introduction}

The launch of the Laser Interferometer Space Antenna (LISA)~\cite{LISA:2017pwj,LISA:2024hlh}, scheduled for the mid-2030s and led by ESA, in collaboration with NASA, will usher in a new era of precision gravitational-wave (GW) astronomy. Operating in the millihertz frequency range—well below that accessible to ground-based detectors such as LIGO, Virgo, and KAGRA—LISA will be sensitive to a wide variety of sources. These include massive black hole binaries (MBHBs)~\cite{Klein:2015hvg,Barausse:2023yrx}, extreme mass ratio inspirals (EMRIs)~\cite{Babak:2017tow}, stellar-mass black hole binaries (stBHBs)~\cite{Sesana:2016ljz}, and an exceptionally large population of double white dwarfs (DWDs)~\cite{Korol:2017qcx,Korol+2022,Lamberts:2019nyk}. 

In contrast to the transient signals detected by terrestrial observatories, many of LISA's sources will be persistent, remaining in-band throughout the mission's multi-year duration and overlapping in both time and frequency. The resulting data stream will be signal-dominated, with certain frequencies affected by a confusion foreground produced by millions of DWDs, mainly from our own Galaxy~\cite{Korol:2017qcx,Korol+2022,Lamberts:2019nyk}. While this richness of signals carries a huge potential for deepening our understanding of the astrophysics of compact binaries, it also poses significant data analysis challenges. Disentangling and characterizing thousands of individually resolvable sources requires simultaneously fitting for an unknown number of overlapping sources -- a task known as `the global fit'~\cite{Littenberg:2020bxy,Lackeos:2023eub,Littenberg:2023xpl,Deng:2025wgk,Strub:2024kbe, Katz:2024oqg}.

Due to their large number, Galactic DWDs play a crucial role in the LISA global fit program. Not only do they provide a stochastic foreground, but several thousands of them are also expected to be individually resolvable~\cite{Korol:2017qcx,Korol+2022,Lamberts:2019nyk}. From the astrophysical point of view, DWDs are relics of the binary stellar evolution process. Their GW signals encode valuable information about the end stages of stellar life, common envelope evolution, and the structure of the Milky Way~\cite{LISA:2022yao}. Understanding the complete population of DWDs, from resolved to unresolved systems, is therefore central to LISA's scientific program.

The traditional approach to astrophysical inference with LISA is based on the success of the global fit~\cite{Littenberg:2020bxy,Lackeos:2023eub,Littenberg:2023xpl,Deng:2025wgk,Strub:2024kbe, Katz:2024oqg} which aims to simultaneously model all resolvable signals and estimate their parameters.
In doing so, population inference biases must be avoided by accounting for the selection effects (Malmquist bias) that arise when fitting loud resolvable events and simultaneously modeling the unresolvable background \citep{Adams+2012}. Proposed global-fit methods are computationally intensive, involving thousands of parameters and potentially large degeneracies between them. In practice, one often relies on simplifying assumptions and iterative pipelines, whose scalability and accuracy remain a key concern. In the context of population inference, a global fit procedure can only infer population properties after estimating the parameters of the resolvable sources and the foreground. In contrast, a faster, scalable approach based on simulation-based inference (SBI)~\cite{SBIfrontier:Kranmer+2020} can directly estimate population properties without the need for source-level inference.

The global fit pipeline is typically optimized for source-by-source detection and cataloging~\cite{Johnson:2025oyu}. Although this enables detailed follow-up of high signal-to-noise sources, it introduces limitations for population-level inference. In a traditional framework, the parameters of the detected sources are first inferred and used to constrain population parameters, such as those that describe the mass function, frequency distribution, or spatial density~\cite{Korol_Rossi_Barausse_2018} (see also \cite{Adams:2012qw,Delfavero_Breivik+2025}). This two-step approach is vulnerable to selection biases, particularly in confusion-dominated regimes, and discards potentially valuable information from sources that fall below the detection threshold (i.e., from the foreground, which also contains information about population parameters~\cite{Breivik+2020}). Alternatively,  performing both source and population inference simultaneously within the global fit can be computationally expensive. In contrast, our method estimates the population properties without the need for source-level inference.

In this work, we propose a fundamentally different strategy: to bypass source-level reconstruction altogether and perform population inference directly from the LISA time (or frequency) series. Our method is based on SBI~\cite{SBIfrontier:Kranmer+2020}—also known as likelihood-free inference—which has emerged in recent years as a powerful alternative to traditional methods in scenarios where the likelihood is intractable or expensive to compute, but forward simulations are available. We apply SBI to the problem of Galactic DWDs, aiming to extract astrophysical population parameters directly from the raw LISA data, without characterizing individual sources. This differs from other neural network-based hierarchical inference methods, which infer population parameters from resolvable GW sources \cite{Leyde+2024, Mold+2025}.

Our approach involves four components. First, using the prescription of \cite{Korol+2022, Korol+2022_code}, we generate forward simulations of the Galactic DWD population using parameterized astrophysical prescriptions for the population's frequency distribution, mass function, and spatial density within the Milky Way. 
We generate GWs, accounting for the LISA response, using the \texttt{GBGPU} implementation of the FastGB waveforms \citep{Katz+2022, katz2022gbgpu, Cornish_Littenberg_2007} and the instrumental noise from the LISA mission requirement document (as implemented in the LDC software~\cite{LDC2022}) to produce realistic synthetic datasets. Second, we compress the data, extracting key features necessary for inference. Third, we train a masked-auto-regressive normalizing flow~\cite{MAF}—a deep neural network architecture that can model complex, high-dimensional probability distributions—to approximate the posterior over population parameters given observed data. Finally, we evaluate the quality of this inference, and apply posterior recalibration~\cite{Crisostomi:2023tle} techniques to ensure unbiased posterior estimates.

Since simulation-based inference techniques can overfit or be miscalibrated, especially when faced with model degeneracies or limited training data \citep{SNLE_2018, Talts+2018}, we assess the statistical coverage of our posteriors for the population parameters across representative test sets, verifying whether true (simulated) parameters lie within the posterior intervals at the nominal credible rate. Where needed, we apply (approximate) Bayesian recalibration~\cite{Crisostomi:2023tle} to correct for under- or over-coverage, thus ensuring improved coverage. These checks are conditional on modeling the data correctly. In the results presented in this paper, that assumption is automatically satisfied because the observations are generated using the same simulator as the training data, but more powerful tests will be required \cite{Model_mispec_SBI_2021, Model_misspec_2023, Model_misspec_2024} to deploy our method on actual LISA data.

A key advantage of our method is that it exploits information from both resolved and unresolved sources simultaneously, making use of the full signal content (suitably compressed). Moreover, once trained on simulated data, the method is amortized, meaning it can produce immediate posterior estimates for multiple datasets without any retraining or generation of extra simulations. This enables astrophysical inference directly from the data stream—without relying on, or waiting for, global-fit results on individually resolved sources. 
Importantly, the method is extensible: it can be easily adapted to incorporate other LISA source classes--including  MBHBs, stBHBs and EMRIs -- and more realistic detector conditions involving non-Gaussian noise, data gaps, or time-dependent systematics~\cite{Dey:2021dem,Baghi:2019eqo,Spadaro:2023muy,Houba:2024ysm,Burke:2025bun}. While our present analysis focuses on Galactic binaries and Gaussian noise for ease of simulation and the purpose of illustrating the approach, the framework we propose lays the foundation for a unified treatment of all types of sources detectable by LISA. For example, in future analyses that include multiple source classes, we can re-use pre-computed simulations of GWs from the DWD population. This ensures that the simulation budget does not scale dramatically, but only increases to the extent necessary to provide adequate coverage of the additional source populations. Several challenges, however, remain to be addressed before this can be achieved, and we discuss them in the Conclusions.

The paper is structured as follows. In Section~\ref{sec:method}, we introduce our pipeline for generating and summarizing LISA data, describe the design and training of the neural network, detail the post-processing calibration of the SBI posteriors, and explain how we obtain inference from the two LISA channels. In Section~\ref{sec:results}, we present our inference outcomes, examine their correlations and parameter-dependent accuracies, and discuss the limitations of our approach. Section~\ref{sec:conclusion} offers a concise summary of the methodology, and outlines avenues for future improvements to our framework. Finally, additional technical details and complementary figures are provided in Appendix~\ref{sec:appendix}.

\section{Method} 
\label{sec:method}
Our aim is to develop an amortized SBI framework to enable rapid posterior evaluation over a wide range of injected population parameter values. Once LISA starts its observation and its real data is available, non-amortized inference strategies, such as sequential SBI \citep{SBIfrontier:Kranmer+2020, TMNRE_Miller+2021}, can yield improved posterior accuracy due to optimized allocation of simulation resources around the observed data. 
In this work, however, we limit ourselves to a faster, amortized, and non-sequential approach, in order to systematically assess the performance of our method across the population parameter space. An extension of our method to focus on a particular data set (i.e., when LISA begins observation) can be easily achieved with a multi-round inference approach.

\begin{figure*}
    \centering
    \includegraphics[width=\linewidth]{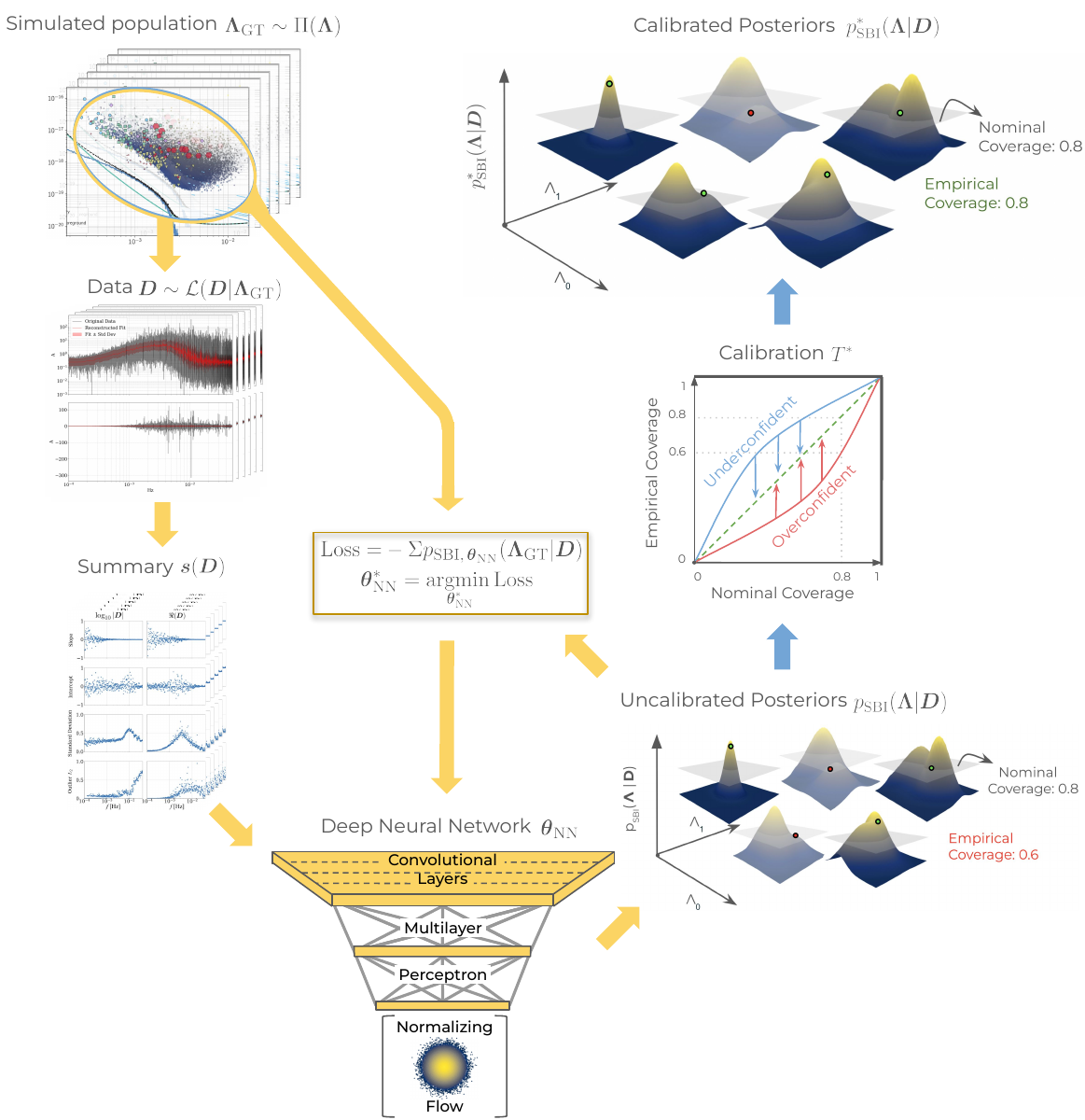}
\caption{Flowchart depicting the SBI training pipeline. The yellow arrows indicate the steps involved in training the deep neural network, and the blue arrows represent the calibration procedure. In the posterior plots, examples of the posterior probability of the injections $p_{\mathrm{SBI}}(\boldsymbol{\Lambda}_\rm{GT} \vert \boldsymbol{D})$ are shown by the green (red) circles representing cases where the injection lies within (outside) the contours of nominal coverage 0.8. The corresponding empirical coverage is marked in the calibration PP plot. The figure illustrating the simulated population is adapted from \cite{LSIA_redbook_2024}.}
    \label{fig:flowchart-training}
\end{figure*}

In this section, we describe the simulations of the DWD population in the Milky Way (\S\ref{sec:method:pop-sim}), the generation of their frequency-domain GW strain (\S\ref{sec:method:data-sim}), the extraction of compressed summary statistics (\S\ref{sec:method:data-summary}), the architecture and training of the deep neural network (\S\ref{sec:method:sbi}), and the calibration in post-processing (\S\ref{sec:method:calibration}). To help illustrate the steps described in this section, Fig.~\ref{fig:flowchart-training} depicts our pipeline from population simulations to calibrated posterior generation. Finally, we detail how we combine the posteriors from the two LISA data channels (\S\ref{sec:AEposteriors}).

\subsection{Data Generation} 
\label{sec:method:data-gen}

As the name implies, SBI requires a dataset of simulations to train a deep neural network, with inference quality and fidelity dependent on the accuracy of the simulator. Therefore, developing a fast, reliable forward simulator that closely models observations is essential \citep{Perigrine_2023}. In the context of inferring the Galactic DWD population from LISA data, the forward simulator must first generate a plausible astrophysical population of the DWDs within the Milky Way, consistent with electromagnetic observations.
Then, the simulator must produce the LISA data corresponding to the astrophysical sources and the expected instrumental noise. For this first attempt at DWD analysis, we ignore the presence of other sources (e.g. MBHBs, stBHBs, and EMRIs) within the LISA data.

\subsubsection{Population model} 
\label{sec:method:pop-sim}

We adopt the Galactic DWD population simulator from the observation-driven model of \cite{Korol+2022}, which is based on the analysis of the spectroscopic samples from the Sloan Digital Sky Survey (SDSS) and the Supernova Ia Progenitor surveY (SPY) from Ref.~\cite{Maoz+2018}. Interestingly, \cite{Korol+2022} finds that their estimates of the DWD population differ significantly from those predicted by binary population synthesis codes. The discrepancy between population synthesis estimates and observation-driven estimates, combined with the long runtime of the former, motivates us to build our forward simulator based on the latter model.
We first summarize the parametric model below and later our GPU-optimized implementation. Our astrophysical model for the DWD population is intentionally simplified (to allow for fast simulations). For instance, we model the primary-mass distribution with a two-parameter Lorentzian. This is suitable for a proof-of-concept, but if the true distribution exhibits structure not captured by this form (e.g., sharp gaps or multimodality), the resulting inference may be systematically biased. 
Extending to more flexible models is straightforward but increases dimensionality. In such cases, dimensionality-reduction techniques (e.g., PCA), as in \cite{PCA_pop_TaylorGerosa2018}, can mitigate complexity while retaining the dominant modes of variation. Alternatively, performing simulations from astrophysical population synthesis models is conceptually possible, but requires a dramatic speed up of the latter (via e.g. emulators).

Ref.~\cite{Korol+2022} parameterizes the DWD population by their mass distribution,  separation distribution, and their spatial number density in the Galaxy. The primary (more massive) star's mass ($m_1$) distribution is modeled after the observed single white-dwarf mass distribution \citep{Kepler+2015}. Based on the observed flat mass ratio distribution \cite{Moe_DiStefano_2017, Duchene_Kraus_2013}, the secondary mass ($m_2$) is sampled from a uniform distribution in the range $[0.15 \Msun, m_1]$. Ref.~\cite{Maoz+2018} finds that the distribution $n_a$ of the separation ($a$) of DWDs as a function of the age of the Milky Way $t_0$ and masses $m_1$, $m_2$ is well approximated by
\be
n_a(x) \propto 
\begin{cases} 
x^{4+\alpha} \left[ \left( 1 + x^{-4} \right)^{\frac{\alpha+1}{4}} - 1 \right] & \text{for } \alpha \neq -1, \\ 
x^3 \ln \left( 1 + x^{-4} \right) & \text{for } \alpha = -1 
\end{cases}
\label{eq:alpha}
\ee
where $x \equiv a/(Kt_0)^{1/4}$ is the normalized separation and
\be
K = \frac{256}{5} \frac{G^3}{c^5} m_1 m_2 (m_1 + m_2).
\ee

Eq.~\eqref{eq:alpha} is approximately a broken power-law with index $\alpha$ when the DWDs merger time is greater than the age of the Milky Way disk ($x \gg 1$). In the opposite limit $x \ll 1$, the power-law index is 3 for $\alpha \geq -1$, and $\alpha + 4$ for $\alpha < -1$. To produce LISA data, the DWD separation is drawn from Eq.~\eqref{eq:alpha}, with a minimum separation of 20,000 km, and a maximum corresponding to the LISA minimum frequency ($\sim 10^{-4}$~Hz). 

The Galactic disk spatial distribution of  DWDs is assumed to follow that of  single white dwarfs, which is modeled as an exponential radial stellar profile with an isothermal vertical distribution given by
\be
n_{\rm d}(R, z) = n_{0} e^{-\frac{R}{\Rd}} \text{sech}^2 \left( \frac{z}{\zd} \right),
\label{eq:wd-dist}
\ee
where $n_{0} = (4.49 \pm 0.38) \times 10^{-3}\,\rm{pc}^{-3}$ is the local white dwarf density from the estimates of \cite{Hollands+2018} from the observational data of \cite{Gaia2018}.
$R$ and $z$ are Galactocentric cylindrical coordinates, while $\Rd$ and $\zd$ are free parameters describing the Milky Way disk's scale width and height. To estimate the distance to the LISA detector, which enters in the waveform, one also needs the position of the Sun in Galactocentric coordinates, ($\mathrm{R}_\odot$, $\mathrm{z}_\odot$), which we fix to (8.1, 0.03)~kpc \cite{Gravity:2019}. 

Besides this disk population, we also consider a Galactic bulge population of DWDs (absent in Ref.~\cite{Korol+2022}). We follow Ref.~\cite{Korol_Rossi_Barausse_2018} and parametrize its density $n_{\rm b}$ as
\be\label{bulge}
n_{\rm b}(r)\propto\, e^{-r^2 / 2\rb^2}\,, 
\ee
where $r=\sqrt{R^2+z^2}$ and $\rb$ is the bulge scale radius. The number of DWDs in the bulge relative to the total number of Galactic DWDs is parametrized by a fraction $\fb$.

Ref.~\cite{Korol+2022} computes the total number of Galactic DWDs, $\N$, as the product of the number of white dwarfs in the Milky Way --i.e., for us, the sum of the integrals of Eqs.~\eqref{eq:wd-dist}--\eqref{bulge} -- and the fraction of white dwarfs in binaries, $f_\rm{DWD}$.In our case, instead of inferring $\N$ from $\rho_{0}$, $\fb$, $f_\rm{DWD}$, $\Rd$, $\zd$ and $\rb$, we leave $\logN$ as a free parameter. This eliminates the need to integrate Eqs.~\eqref{eq:wd-dist}--\eqref{bulge} for each realization of $\Rd$, $\zd$ and $\rb$. Additionally, we simplify the model of the primary mass distribution. Refs.~\cite{Kepler+2015, Korol+2022} model the distribution as a mixture of three Gaussian components, truncated between 0.15 and 1.4$\Msun$, resulting in eight free parameters. Here, we instead model the distribution using a single truncated Lorentzian, characterized by only two free parameters, $\m0$ and $\mgamma$, yet capable of producing qualitatively similar results to the Gaussian mixture. Further details on our choice of mass function are given in the Appendix~\ref{sec:appendix:Mass_fit}, where we compare it to the observed single white-dwarf mass distribution, as well as the Gaussian mixture of \citep{Kepler+2015}.

\begin{table}[ht]
\centering
\begin{tabular}{l >{\raggedright\arraybackslash}p{0.37\linewidth} l l}
\hline
$\boldsymbol{\Lambda}$ & \textbf{Description} & \textbf{Prior} & \textbf{Fiducial} \\
\hline
$\m0$         & Primary-mass peak [$\Msun$]       & $\mathcal{U}(0.15, 1.4)$          & 0.649 \\
$\logmgamma$  & Primary-mass log half-width at half-maximum & $\mathcal{U}(-2.0, 0.5)$   & 2.686 \\
$\Rd$         & Disk scale radius [kpc]           & $\mathcal{U}(1.0, 4.0)$           & 2.5 \\
$\zd$         & Disk scale height [kpc]           & $\mathcal{U}(0.1, 1.0)$           & 0.3 \\
$\rb$         & Bulge scale radius [kpc]          & $\mathcal{U}(0.1, 1.0)$           & 0.5 \\
$\fb$         & Bulge fraction                    & $\mathcal{U}(0.05, 0.6)$          & 0.27 \\
$\alpha$      & DWD separation index              & $\mathcal{U}(-2.05, -0.55)$       & -1.3 \\
$\logN$       & DWD log-number                    & $\mathcal{U}(6.3, 7.8)$           & 6.8 \\
$\A$          & Acceleration noise factor         & $\mathcal{N}(1, 0.2)$             & 1.0 \\
$\O$          & Optical noise factor              & $\mathcal{N}(1, 0.2)$             & 1.0 \\
\hline
\end{tabular}
\caption{Model parameters $\boldsymbol{\Lambda}$ and their prior distribution, its range, and fiducial values. $\mathcal{U}(a,b)$ refer to the uniform distributions with limits $a$ and $b$. $\mathcal{N}(\mu, \sigma)$ refers to the one-dimensional normal distribution with mean $\mu$ and standard deviation $\sigma$.}
\label{tab:parameters}
\end{table}

We adopt a uniform prior over our population parameters $ \{\m0,\,\logmgamma,\, \Rd,\, \zd,\, \rb,\, \fb,\, \alpha,\, \logN\}$. We set astrophysically agnostic prior limits on all parameters, such that they contain at least the 5-$\sigma$ observational electromagnetic constraints. $\m0$ and $\logmgamma$, the primary-mass distribution parameters, have values ranging from [0.15, 1.4]~$\Msun$ and [-2, -0.5] respectively. The Galactic parameters $\Rd$, $\zd$, $\rb$, $\fb$ have ranges [1.0, 4.0]~kpc, [0.1, 1.0]~kpc, [0.1, 1.0]~kpc, and [0.05, 0.6]  respectively based on \cite{MWdisclength_2008, MWscaleheight_2017, MWscalebulge_2009}. $\alpha$ has range [-2.05, -0.55] \citep{Maoz+2018} and $\logN$ varies from [6.3, 7.8] based on the 5-$\sigma$ uncertainties on $\rho_0$ \citep{Hollands+2018} and $f_\rm{DWD}$ \citep{Korol+2022}. We also define a fiducial value based on those considered in \cite{Korol+2022, Korol_Rossi_Barausse_2018}. Table~\ref{tab:parameters} summarizes the population parameters, their descriptions, prior distributions, and fiducial values.

For our analysis, we adopt the codes of \cite{Korol+2022_code, natalia_code_2022} based on the model from \cite{Korol+2022}, but with our additional Milky Way bulge population and simplified parametric mass distribution. Our novel GPU-accelerated implementation\footnote{Code to be released shortly.} that leverages the computing facilities of CuPy and Numba, a just-in-time (JIT) compiler, yields three orders of magnitude faster population generation than \cite{Korol+2022_code}. For a given set of parameter values, the code can produce a Galactic realization in about a minute. This enables us to simulate millions of possible DWD populations, each with tens of millions of DWDs.

\begin{figure}
    \centering
    \includegraphics[width=\linewidth]{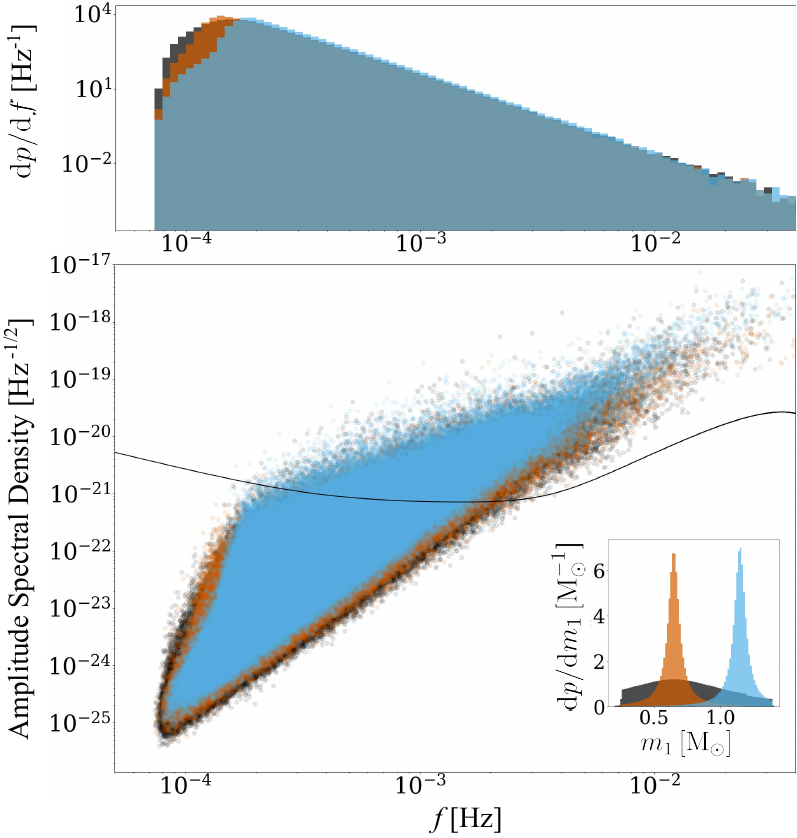}
    \caption{Top: probability density of double white dwarfs (DWDs) as a function of GW frequency. Bottom: frequency spectrum of DWD populations with different primary mass $m_1$ distribution (inset). The black curve shows the frequency response of the LISA A-channel. The different colors are for different primary-mass distributions (and all other parameters fixed at fiducial values): $\m0, \mgamma$ = [0.65, 0.0485], [1.25, 0.0485], [0.65, 0.31] (orange, blue, gray respectively). Note that this is the amplitude spectral density of the channel, and not the strain amplitude. }
    \label{fig:pop}
\end{figure}

Fig.~\ref{fig:pop} illustrates the frequency distribution of three realizations of the DWD population from different primary-mass distributions and other parameters fixed at a fiducial value.
The inset shows the absolute value of the frequency spectrum generated for the three DWD populations, differing only in their primary mass distribution. We overplot the LISA typical instrumental noise amplitude spectral density (ASD) for reference. 
% , assuming here $A=O=1$.  
The vast number of overlapping sources at low frequencies ($10^{-4}-10^{-3}$~Hz) contributes to the Galactic ``foreground'' or ``confusion noise'', an unfortunate term considering that some of the parameter inference signature could potentially lie here. We explore this in further detail in \S~\ref{sec:discussion:WhereInfo}. 

\subsubsection{Noise Model and Data Simulation} 
\label{sec:method:data-sim}
We consider 4 years of LISA observations at a cadence of 10 seconds. We assume stationary Gaussian instrumental noise and ignore the presence of glitches and other possible noise artifacts.  We use the noise power spectral density (PSD) from  the LISA mission requirements document MRDv1 as implemented in~\cite{LDC2022}. 
Generating realistic time-varying non-Gaussian noise currently requires end-to-end instrument simulators \citep{LISAInstruments_2023}, which are currently computationally expensive for hundreds of thousands of multi-year, 10-s-cadence runs. We therefore adopt a PSD-driven stationary noise. This is a pragmatic choice, and not a limitation of the method, because the pipeline can be re-trained with more realistic, non-stationary noise, along with other artifacts, as scalable simulators mature. However, the normalizations of the acceleration noise and the optical metrology system (OMS) noise are assumed to be known only up to factors of order unity. 
\begin{equation}
S_{n}(f) = \A S_{\text{acc}}(f) + \O S_{\text{OMS}}(f)\,.
\end{equation}
\label{eq:noise}
The parameters $\A$ and $\O$ are part of our model and are inferred from the data, assuming a Gaussian prior, i.e.
\begin{equation}
p(\A, \O) \propto \exp\left( -\frac{1}{2} \left[ \frac{(\A - 1)^2}{\sigma_\A^2} + \frac{(\O - 1)^2}{\sigma_\O^2} \right] \right)\,,
\end{equation}
with $\sigma_\A=\sigma_\O=0.2$. As shown in Table~\ref{tab:parameters}, the fiducial value of $\A$ and $\O$ is 1.

We produce the frequency domain waveform of each DWD using the GPU-accelerated software package GBGPU \citep{Katz+2022, katz2022gbgpu, Cornish_Littenberg_2017, GBGPU-ref4}. 
Given our LISA observation scenario, the GBGPU package provides the waveform for each DWD at 128 frequencies centered about the dominant frequency. The primary bottleneck is to sum the contributions of the tens of millions of DWDs in a given population into a single frequency series of length $\sim 6 \times 10^6$. The typical vectorized broadcasting method fails due to the size of the data. 
We instead leverage atomic addition within the GPU-accelerated JIT compiler, Numba. When possible, we keep most of the data generation computation on GPU, minimizing the time-consuming data transfer between CPU and GPU. To this end, for each data channel, we use Numba to sum the tens of millions of DWD GW strains, each occupying 128 frequency bins, onto the noise spectrum realization. 

Finally, as a pre-processing step for the SBI, we whiten the data by dividing out a fiducial instrumental noise PSD given by Eq~\ref{eq:noise}, setting $\A=\O=1$.\footnote{ Note that 
the noise parameters $\A,\O$ {\textit are}
sampled in our approach. The choice $\A=\O=1$ is only adopted for the data whitening, which must be the same for all simulated/observed data.}

 The total time to generate a single data realization from population to strain generation is $\sim 55$~s, as compared to a few hours before our accelerated implementation. Each realization contains $\sim 6 \times 10^6$ complex 64-bit floating-point values occupying 80MB of memory. This makes it impractical to load multiple realizations in batches into the limited volatile memory of most GPUs for training the SBI. Thus, it is necessary to compress and summarize the data. Alternatively, one would need to construct an optimal embedding representation that encapsulates the context of the data.

In summary, we simulate data from the 10-dimensional model parameter $\boldsymbol{\Lambda}\equiv [\m0,\,\logmgamma,\, \Rd,\, \zd,\, \rb,\, \fb,\, \alpha,\, \logN, \, \A, \, \O]$ whose prior distribution $\Pi(\boldsymbol{\Lambda})$ is uniform for the eight population parameters and a standard normal for the two noise parameters. We denote the simulated data by $\boldsymbol{D}_\mathrm{A}$ and $\boldsymbol{D}_\mathrm{E}$ for the two independent LISA channels A and E, respectively. When referring to data from either channel, we use $\boldsymbol{D}$ to denote a generic dataset from A or E.

\subsection{Compressed Data Summary} 
\label{sec:method:data-summary}
Data summaries, $\boldsymbol{s}(\boldsymbol{D})$, are compressed representations of high-dimensional data, $\boldsymbol{D}$, that ideally retain most of the information needed to estimate the parameters of the model $\boldsymbol{\Lambda}$ 
\citep{IMNN_2018, SBIfrontier:Kranmer+2020, RobustSummaries_ModelMispec_2023}, while greatly reducing data dimensionality, $\dim(\boldsymbol{s}(\boldsymbol{D})) \ll \dim(\boldsymbol{D})$. An ideal lossless compressed summary would lead to the same posterior estimate as the full data, $p(\boldsymbol{\Lambda} \vert \boldsymbol{s}(\boldsymbol{D})) = p(\boldsymbol{\Lambda} \vert \boldsymbol{D})$. The reduced dimensionality enables larger batches of data to be loaded into memory, resulting in faster SBI training -- albeit potentially incurring some information loss. 

We explored conventional dimensionality reduction techniques, including Principal Component Analysis (PCA) and autoencoders, to compress high-dimensional frequency series data into informative summaries for the SBI. Although these methods are widely used and effective for capturing dominant modes of variation \cite{SBIfrontier:Kranmer+2020}, they can be difficult to train when compressing large data, as in the case of years-long LISA data.
Indeed, we find that naive implementations of autoencoders and PCAs tend to focus on certain noise features (such as particularly large noise fluctuations). For PCA, achieving 95\% cumulative explained variance requires more than $10^{4}$ components; using fewer components fails to recover the peak amplitudes and the overall power–spectrum shape of resolvable sources. For autoencoders, the parameter count grows rapidly with network width and depth, so the trainable encoder–decoder capacity is constrained by GPU memory, which limits reconstruction fidelity. To address these limitations, we developed a compression algorithm tailored to the specific characteristics of the DWD population spectrum. We design our summary to retain physically meaningful features of the bump in the signal power spectrum from the tens of millions of unresolved DWDs, while also providing statistical information on the few loud, resolvable sources. 

First, we simplify the complex 64-bit vector representation by concatenating the 32-bit log-absolute and real values of the data, hereon denoted as $\log_{10}\vert\boldsymbol{D}\vert$ and $\Re(\boldsymbol{D})$, respectively. We empirically find that this representation offers slightly more robust training in comparison to other variations, including the real-imaginary and absolute-phase tangent pairs, as well as just the real, imaginary, or absolute values. Interestingly, variants incorporating absolute values exhibit marked improvements in the posteriors of SBI compared to those that do not. This suggests that the underlying signal-to-noise ratio of the sources may carry more informative content for population inference than the phase information associated with individually resolvable sources. Alternatively, as our summaries are binned, the averaging may cause us to lose phase information while retaining amplitude. We leave this for future investigation.
However, combining the absolute value with either the real or imaginary component yields the most effective posterior estimates, offering improved coverage and accuracy (as further elaborated in \S~\ref{sec:results}). This suggests that jointly leveraging both amplitude and partial phase information may capture complementary aspects of the signal relevant to inference. 

We apply the following compression procedure independently to each channel's
frequency series realization $\boldsymbol{D}$. We first rebin the original $6 \times 10^{6}$-dimensional frequency series by averaging over non-overlapping windows of 10 consecutive frequencies, resulting in a vector of size $6 \times 10^{5}$. The averaged data are further subdivided into 1024 logarithmically spaced frequency bins. We also explored larger (2048) and smaller (256) numbers of bins, as well as linear bin spacing, but found log-spaced 1024 bins a good compromise of data compression and expressivity. Within each bin, we linearly fit the values as a function of frequency, with the bin-wise slope and intercept as free parameters. We then compute the standard deviation of the residuals from this linear fit, $\pm\sigma_\mathrm{i}$, for each bin of index $i$. We identify residual outliers as those that lie beyond $\pm3\sigma_\mathrm{i}$, and quantify their importance using their $L_2$ norm, calculated as the square root of the sum of their squared residuals. These outliers are likely due to loud resolvable sources, and their $L_2$-norm thus serves as a metric that roughly quantifies their contribution to the data. Hence, for each frequency bin the data summary has four features: the slope, the intercept, the standard deviation of the residuals, and the $L_2$-norm of the residual outliers. 

Note that a linear fit of the absolute value of the frequency series resulted in residuals skewed toward negative values, i.e., the fit consistently overestimated the amplitude. However, a linear fit of the log-absolute value resulted in less biased, zero-mean residuals, suggesting that the log-amplitude may be better suited for our compression algorithm. The linear fit of the real values showed no significant biases in the residuals. Fig.~\ref{fig:residual} in Appendix~\ref{sec:appendix:Residual_distribution} illustrates the distributions of the residuals for a few example frequency bins (c.f. legend) and all the bins (inset). The generally symmetric and zero-mean distribution of the residuals indicates that the standard deviation can be used as a rough metric to quantify the variation within each frequency bin. The $L_2$-norm of the residual outliers summarizes significant deviations not described by the standard deviation.

\begin{figure}
    \centering
    \includegraphics[width=\linewidth]{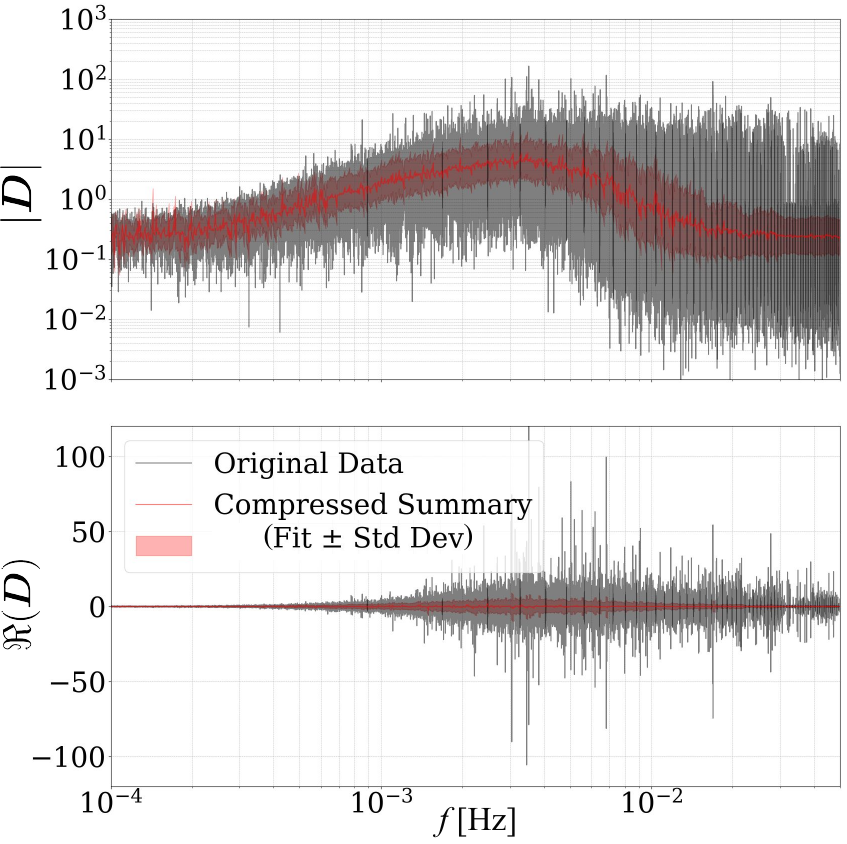}
    \caption{Comparison of the log-absolute $\log_{10}\vert\boldsymbol{D}\vert$ (top) and real $\Re(\boldsymbol{D})$ (bottom) values of the $6\times10^{6}$ dimensional data (black) that are summarized by a piece-wise linear fit in 1024 frequency bins (red curve) and the $1\sigma$ residual contour (red shade).}
    \label{fig:data}
\end{figure}

\begin{figure}
    \centering
    \includegraphics[width=.9\linewidth]{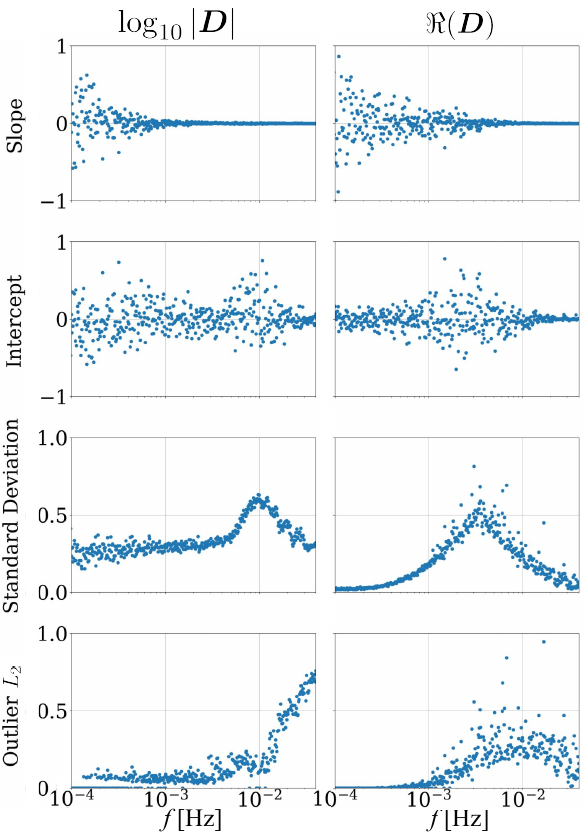}
    
    \caption{Comparison of the $4\times1024$-dimensional data summaries for $\log_{10}\vert\boldsymbol{D}\vert$ and $\Re(\boldsymbol{D})$. Each frequency bin is summarized by four features: the slope and intercept of the linear fit, the standard deviation of the residual, and the $L_2$ norm of the residual outliers. Note that each feature has been scaled to a comparable range for stable training.
    }
    \label{fig:datasummary}
\end{figure}

The nearly $6 \times 10^{6}$ dimensional complex 64-bit data $\boldsymbol{D}$ is therefore compressed into two $4096$-dimensional 32-bit summaries (i.e., four features for each 1024 frequency bin), $\boldsymbol{s}(\log_{10}\vert\boldsymbol{D}\vert)$ and $\boldsymbol{s}(\Re(\boldsymbol{D}))$, corresponding to a compression factor of $\sim1.5\times10^3$. Fig.~\ref{fig:data} illustrates an example of the log-absolute and real components of the high-dimensional original data as well as the compressed description from the fit + residual standard deviation. Finally, we normalize each feature at every frequency bin to ensure they are on a consistent scale. 
The corresponding data summary values are shown in Fig.~\ref{fig:datasummary}.

In comparison to generic compression methods (such as PCA, autoencoders, etc.), we find empirically that this summary delivers improved training performance in the downstream inference task. We leave the exploration of more advanced autoencoder models and higher-order functional approximations for future improvements of the algorithm.

\subsection{Deep Neural Network} 
\label{sec:method:sbi}
We train two separate neural networks to compute posteriors from the LISA A and E channels, respectively. Our deep neural network implementation comprises two distinct sections. The first component, the \textit{context encoder}, extracts relevant correlations of the data summary onto a lower-dimensional embedding space. The second component is a normalizing flow, which computes the posterior probability density based on the context provided by the encoder. We pass, as input to the network, the data summary of length 1024 with eight features corresponding to the four summaries per component of data (log-absolute and real). 

We use a typical convolutional neural network architecture that passes the data summary through four successive convolutional blocks and a final multi-layer perception. Each block contains a fixed kernel size (8), stride (2), and padding (3). However, the number of channels progressively increases from 8 to 256 across the four blocks. Starting with the input 8 channels of length 1024 ($8\times1024$), successive convolutional blocks have dimensions $32\times512$, $64\times256$, $128\times128$, and $256\times64$. 

As the spatial resolution of the feature maps decreases due to striding (by a factor $\sim2$ after each block), increasing the channel dimension (also by a factor 2) ensures that the network retains sufficient representational capacity. After the convolutional blocks, the output is flattened and passed through two fully connected layers (each of size 256 units). Batch normalization is applied after each convolutional layer to mitigate internal covariate shift and stabilize training, while dropout is used after the convolutional stack and each fully connected layer to reduce overfitting.

The second component is a conditional normalizing flow to model the posterior probability density. We implement a Masked Autoregressive Flow (MAF) \citep{MAF} via the nflows library \citep{nflows}. The data embeddings from the context encoder are used to condition the flow's transformation of a ten-dimensional standard normal distribution into the ten-dimensional posterior density $p_\rm{SBI}(\boldsymbol{\Lambda}\vert\boldsymbol{s}(\boldsymbol{D}))$. We design a MAF with 16 transformations, each with 4 residual blocks with 512 hidden units. Within each residual block, densely connected layers parameterize the shift and scale functions required by the autoregressive transformation. 

The deep network is trained by a cross-entropy loss that maximizes the average predicted posterior for the training dataset. We illustrate the training procedure by the yellow arrows in Fig.~\ref{fig:flowchart-training}, and we summarize the architecture of the deep neural network in Tables \ref{tab:CNN}, \ref{tab:NF} of \S~\ref{sec:appendix:nn}. We train the network on a dataset comprising $3 \times 10^6$ pairs of $\{ \boldsymbol{\Lambda},  \boldsymbol{D}(\boldsymbol{\Lambda})\}$. 
$\boldsymbol{\Lambda}$ is sampled from a prior distribution $\Pi(\boldsymbol{\Lambda})$ that is uniform over all parameters except for the noise, which follows a Gaussian distribution as specified in \S\ref{sec:method:pop-sim} and \S\ref{sec:method:data-sim}. For each sampled $\boldsymbol{\Lambda}$, we generate ten independent realizations of the DWD population, and for each population realization, we generate ten LISA instrumental noise realizations. This approach enhances training efficiency, particularly by improving the model's ability to distinguish between noise and signal for a given parameter. We employ a training strategy that reduces the learning rate by half if the validation loss does not decrease for more than three epochs and stops training if the loss does not improve by at least 10\% over ten epochs. It takes 8 hours on an NVIDIA A100 GPU to train the network for a given channel. Inference of the posterior distribution for a given realization is produced in a few seconds.

\subsection{Coverage \& Calibration} 
\label{sec:method:calibration}

An ideal SBI network should provide posterior probability distributions that faithfully represent posterior uncertainty, for example, by exhibiting exact coverage.  In practice, due to the limited number of simulations, limited expressivity, and data compression, there can be a discrepancy in the approximation obtained by the network of such an ideal posterior

A credible region with nominal coverage $c$ is defined as the posterior region that contains the true parameter value with posterior probability $c$. In our case, we use the highest posterior density (HPD) region, which ensures that the credible region is the smallest possible set with this probability. The empirical coverage quantifies how often the true parameters actually fall within the specified (i.e., nominal) credible regions. If the posteriors are well-calibrated, the empirical and nominal coverages should match.

A probability–probability (PP) plot shows the empirical coverage against the nominal credibility levels. Ideally, a calibrated inference method yields a PP plot that lies along the diagonal line (the line of perfect calibration), indicating perfect alignment between expected (nominal) and actual (empirical) coverage. Deviations from this diagonal indicate under-confidence (curve above the diagonal) or over-confidence (curve below the diagonal) in the posterior distributions. We illustrate this diagrammatically in the PP plot of Fig.~\ref{fig:flowchart-training}. Note that in a Bayesian PP plot (i.e., one that is averaged over the prior distribution for the parameters), the diagonal line is a necessary, but not sufficient, condition for exact coverage: some regions in the parameter space might overcover (above the $45^\circ$ line), while others under-cover (below the $45^\circ$ line), achieving exact coverage on average but not necessarily anywhere in parameter space. Since amortized SBI enables fast posterior computation, it becomes feasible to efficiently evaluate PP distributions and thereby rapidly assess the quality of the inferred posteriors. This would be computationally expensive and time-consuming to compute for traditional likelihood-based Bayesian samplers. 

\begin{figure}
    \centering
    \includegraphics[width=\linewidth]{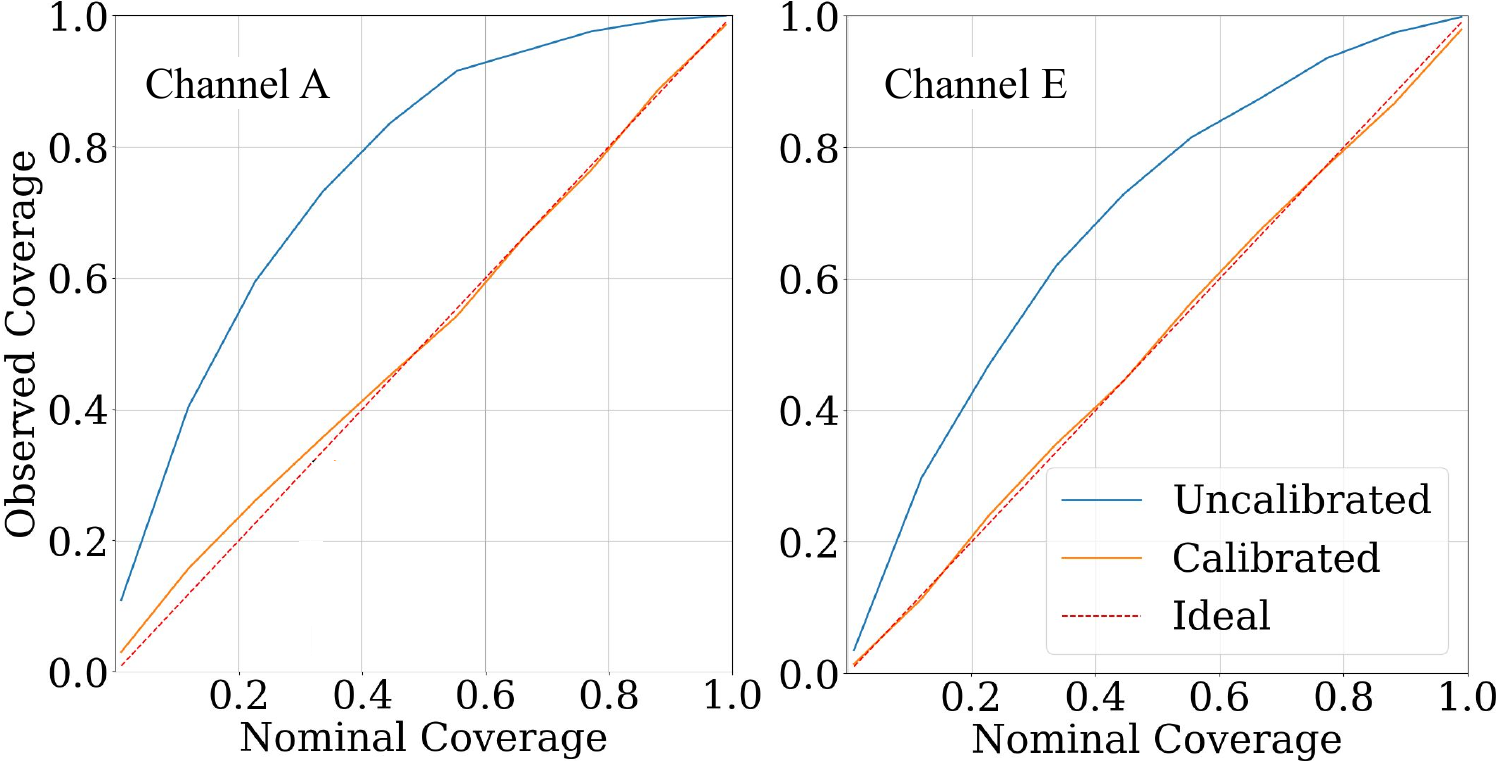}
    \caption{Empirical (i.e., observed) vs nominal coverage for the ten-dimensional posterior before (blue) and after  (orange) calibration for the SBI trained on channel A (left) and E (right) data. The black dotted $45^\circ$ line shows the ideal relation. The optimal calibration temperatures $T^*$ for the two channels are 0.62 and 0.72, respectively.}
    \label{fig:coverage}
\end{figure}

We evaluate the credible regions of the ten-dimensional posteriors by first computing the inverse cumulative distribution function ($F^{-1}$) from posterior samples, sorted in descending order of probability density. For a given credibility level $c_0$, the corresponding nominal coverage is defined as the region of parameter space containing all samples with cumulative posterior mass $c \leq c_0$. As the cumulative mass is calculated starting from the probability density maximum, this region is known as the highest-probability density (HPD) region for that credibility level $c_0$. Mathematically, this is a region of $\boldsymbol{\Lambda}$ satisfying $p(\boldsymbol{\Lambda}\vert \boldsymbol{D}) \geq $ $p(F^{-1}$($c_0$)). An HPD region has the property of being the shortest interval (in one dimension) containing a fraction $c_0$ of posterior probability.

If the PP curve does not follow the ideal $45^\circ$ line, the SBI samples can be calibrated in postprocessing. To this end, we calibrate the posteriors with an optimal temperature $T^*$ by the following expression,

\be
p^*_{\mathrm{\,SBI}}(\boldsymbol{\Lambda} \vert \boldsymbol{D}) = p_{\mathrm{\,SBI}}(\boldsymbol{\Lambda} \vert \boldsymbol{D}) ^ {1/T^*},
\ee
where $p^*_{\mathrm{\,SBI}}(\boldsymbol{\Lambda} \vert \boldsymbol{D})$ and $p_{\mathrm{\,SBI}}(\boldsymbol{\Lambda} \vert \boldsymbol{D})$ are the calibrated and uncalibrated SBI posteriors respectively. $T^*$ is a scalar optimal temperature which minimizes the mean square difference between the ideal PP plot ($45^\circ$ line) and the PP plot computed by scaling the posteriors. If the uncalibrated coverage is under-confident (below the $45^\circ$ line), the SBI posteriors are too wide and require a calibration temperature $T^*<1$ to contract the distributions. Conversely, if the SBI is over-confident (above the $45^\circ$ line), the posteriors are too narrow and require $T^*>1$ to expand the distributions. A perfectly calibrated SBI (on the $45^\circ$ line) corresponds to $T^* = 1$.

In Fig.~\ref{fig:flowchart-training}, we illustrate the calculation and calibration of the empirical coverage for an example nominal coverage of 0.8. Before calibration, we illustrate an empirical coverage of 0.6, where in 3 out of 5 evaluations of the posterior probability, the ground truth $\boldsymbol{\Lambda}_\rm{GT}$ lies above the 0.8 nominal coverage contour. Upon calibrating the overconfident SBI posteriors by a temperature factor, the empirical coverage converges to 0.8 as expected. The blue and orange curves of Fig.~\ref{fig:coverage} show the uncalibrated and calibrated coverages for our SBI trained on the two channels A (left) and E (right).  The uncalibrated coverage for the two SBIs fall above the ideal coverage, indicating under-confidence and hence requiring an optimal temperature $T^*<1$ (0.62 and 0.72, respectively). We provide an example comparing the calibrated and uncalibrated posteriors for the A channel SBI in $\S$~\ref{sec:appendix:CalibrationComparison}. Note that after this procedure, the posteriors are calibrated {\em on average} across the prior distribution -- i.e., there is no guarantee that each individual posterior will exhibit exact coverage (see e.g. \cite{Karchev:2022xyn} for an example of so-called frequentist calibration, which corrects coverage at each parameter value).

\subsection{Combining posteriors}
\label{sec:AEposteriors}
Assuming that the A and E channels of LISA can be treated as independent, and given the calibrated individual channel posteriors $p(\boldsymbol{\Lambda} \vert \boldsymbol{D}_{\mathrm{A}}), \, p(\boldsymbol{\Lambda} \vert \boldsymbol{D}_{\mathrm{E}}) $, the combined posterior can be derived as

\begin{align}
p(\boldsymbol{\Lambda} \vert \boldsymbol{D}_{\mathrm{A}}, \boldsymbol{D}_{\mathrm{E}}) 
&= \frac{\mathcal{L}(\boldsymbol{D}_{\mathrm{A}}, \boldsymbol{D}_{\mathrm{E}} \vert \boldsymbol{\Lambda}) \cdot \mathrm{\Pi}(\boldsymbol{\Lambda})} {p(\boldsymbol{D}_{\mathrm{A}}, \boldsymbol{D}_{\mathrm{E}})} \notag \\
&\propto \mathcal{L}(\boldsymbol{D}_{\mathrm{A}} \vert \boldsymbol{\Lambda}) \cdot \mathcal{L}(\boldsymbol{D}_{\mathrm{E}} \vert \boldsymbol{\Lambda}) \cdot \mathrm{\Pi}(\boldsymbol{\Lambda}) \notag \\
&\propto \frac{p(\boldsymbol{\Lambda} \vert \boldsymbol{D}_{\mathrm{A}}) \cdot p(\boldsymbol{\Lambda} \vert \boldsymbol{D}_{\mathrm{E}})}{\mathrm{\Pi}(\boldsymbol{\Lambda})},
\end{align}
where $\mathrm{\Pi}$ represents the prior, $\mathcal{L}$ the likelihood, and $p(\boldsymbol{D}_\rm{A}, \boldsymbol{D}_\rm{E})$ is the evidence of the data (a constant). In practice, this is equivalent to using the posterior from the E channel inference as the prior for the A channel inference (second equation above). However, since we do not have access to an explicit form for the likelihood, we produce samples from the two-channels posterior by re-weighting posterior samples from channel A by the ratio of their posterior probability from channel E and the prior density, as shown by the last equation. In future iterations of the method, we may explore the evaluation of $p(\boldsymbol{\Lambda} \vert \boldsymbol{D}_{\mathrm{A}}, \boldsymbol{D}_{\mathrm{E}})$ from an SBI with the joint data as input.

\section{Results}
\label{sec:results}

\begin{figure*}
    \centering
    \includegraphics[width=\linewidth]{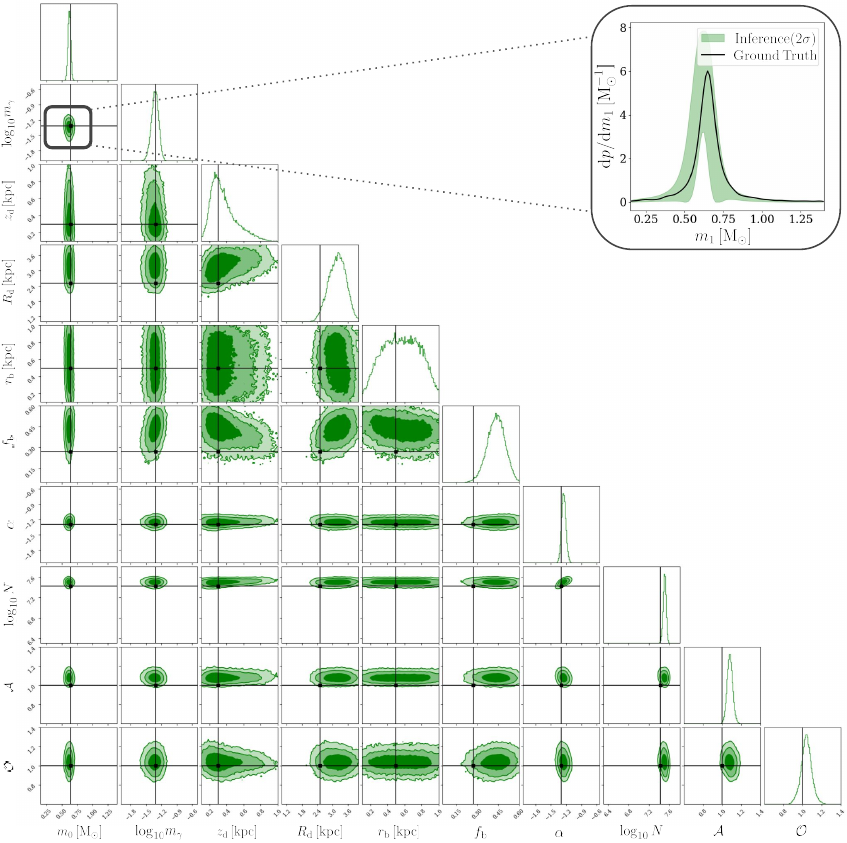}
    \caption{Corner plot of the joint (A and E channels') calibrated SBI posterior distribution. The inset shows the inferred distribution with 2$\sigma$ confidence of the primary mass obtained from the posteriors of the corresponding mass parameters $\m0$, $\mgamma$.  The injected ground truth is indicated in black. The range of the corner tiles represents the amortized prior range of the SBI. }
    \label{fig:posterior}
\end{figure*}

In Fig.~\ref{fig:posterior}, we show the calibrated SBI posteriors for the observation-driven DWD population model of \cite{Korol+2022}. The inset illustrates the primary mass distribution, reconstructed using the posterior samples for $\m0, \, \logmgamma$ in Eq.~\eqref{eq:mass-fit}. The range of values shown in Fig.~\ref{fig:posterior} and all subsequent posterior figures represents the prior box. The best inferred information consists of the primary mass distribution $\rm{d}p\rm{/d}m_1$, the separation distribution index $\alpha$, and the total number of Galactic DWDs $\logN$. In contrast, the posterior contours of the Milky Way's disk scale lengths are less constraining. The mild bias in the recovery of $\A$ is not systematic throughout the SBI, and other realizations and examples result in slightly different posteriors.

Information on the Galactic scale dimensions comes from the sky localization of the sources, whose signature resides in the phase. Since the incoherent, unresolvable sources contribute limited insight into the phase and hence spatial distribution of DWDs, the SBI instead targets the signature of resolvable sources, where the phase information is more accessible. For example, constraints on the Galactic scale height $\zd$ have been computed for the resolvable \cite{Korol_Rossi_Barausse_2018} and unresolvable sources  \citep{Breivik+2020}. Indeed, when passing just the log-absolute data to the summary (where the phase information is lost), the SBI failed to produce meaningful posteriors for the Galactic scale dimensions, approximately reproducing the prior. This was a primary motivation to also include the real component of the data in the summary. We further characterize information from resolvable and unresolvable sources in \S~\ref{sec:discussion:WhereInfo}. Fig.~\ref{fig:posterior} shows that our posteriors on $\zd$ agree with the analysis of resolvable sources in \citep{Korol_Rossi_Barausse_2018}. In comparison to the Galactic disk scale radius $R_d$, the bulge scale radius $r_b$ often has a less constrained posterior, likely due to the distant location of these sources in comparison to those in the disk near the solar system. On a similar note, the bulge fraction is therefore not as well constrained.

\begin{figure}
    \centering
    \includegraphics[width=\linewidth]{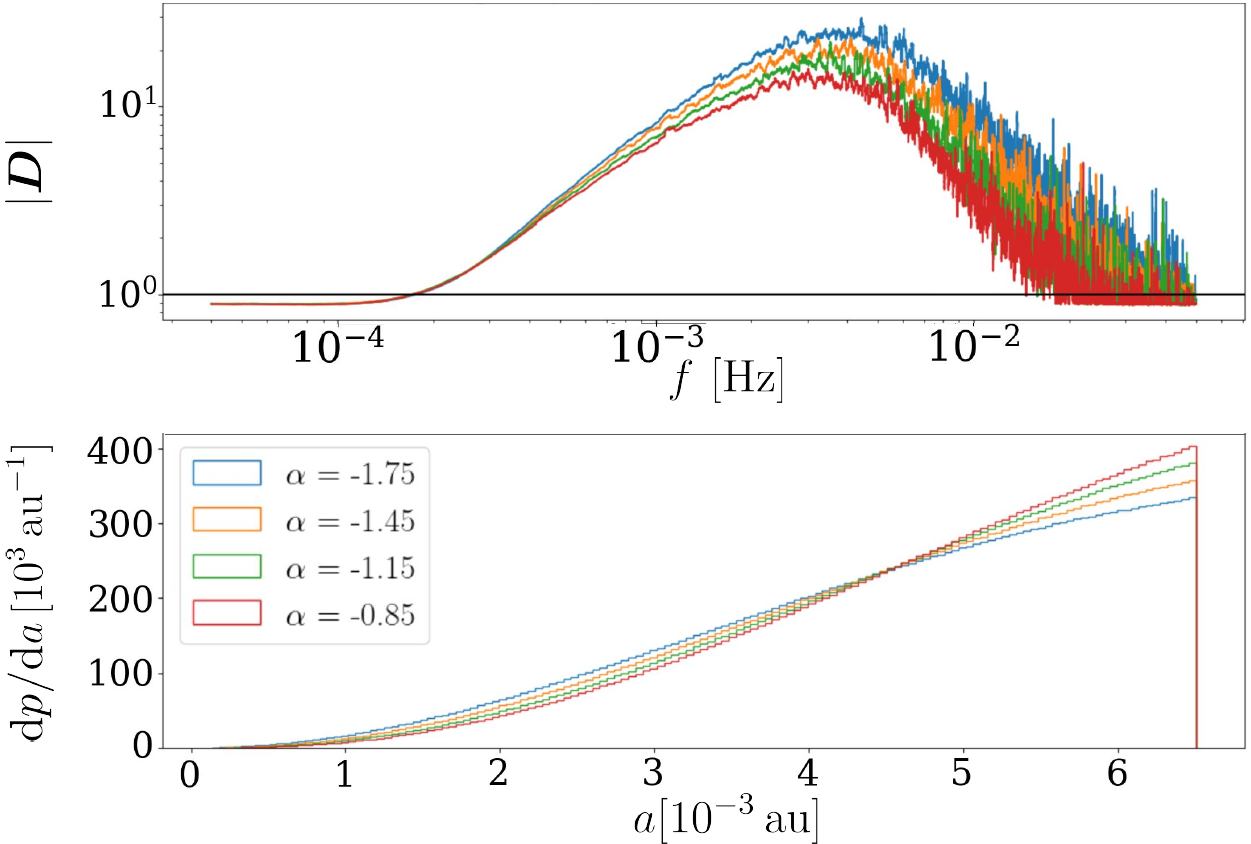}
    \caption{Effect of varying $\alpha$ on the population signal. Increasing $\alpha$ reduces the overall amplitude, while decreasing it enhances the signal strength. For clarity, only the smoothed, noise-free data are shown.}
    \label{fig:alpha_trends}
\end{figure}

The interplay between the distribution of the DWD GW frequencies in relation with the instrumental noise PSD largely determines the relatively confident posteriors and their correlations seen in $\m0$, $\mgamma$, $\alpha$, and $\logN$. Among the well-constrained parameters, the inferences on $\alpha$ and $\N$ are especially narrow and show a positive correlation (seen in Fig.~\ref{fig:posterior}). Decreasing $\N$ and increasing $\alpha$ have a similar signature on the data: both cause a drop in the power spectrum of the unresolved sources. The former is simply because the signal of the population is proportional to the number of sources. Less obvious is the effect of $\alpha$. Increasing $\alpha$ increases the number of sources with higher separation, and thus with a lower GW frequency. Due to the shape of the LISA PSD (black curve in Fig.~\ref{fig:pop}), lower frequencies ($\lesssim 10^{-3}$~Hz) generally result in lower signal-to-noise ratios (SNRs). Fig.~\ref{fig:alpha_trends} illustrates this effect by comparing the smoothened whitened data for different values of $\alpha$.

SBI's ability to perform hierarchical inference on LISA data without the need for explicit likelihood evaluations opens novel avenues of testing and refining complex binary stellar evolution models that predict the formation of DWDs. For example, based on inferences of the DWD primary mass and separation distribution, one may place constraints on stable and unstable mass transfer mechanisms. Such constraints can even help better predict the merger rates of stellar-mass binary black holes originating from isolated binary evolution \cite{Srinivasan+2023, Iorio+2023}. A global fit of the resolvable DWDs in LISA has already been shown to help constrain these models \cite{CE_DWDseperation_Korol+2022, Delfavero_Breivik+2025}. A population-level inference drawing information from orders of magnitude more sources (resolvable and unresolvable) can potentially provide more robust, statistically significant constraints and is an avenue worth exploring in future studies.

These results demonstrate that our amortized SBI can exploit the full DWD signal, resolved and unresolved, to perform population‐level inference on millions of binaries in a single, unified analysis, unbiased by selection effects. The training of the SBI was made feasible by employing a fast forward-simulator, which significantly reduced the data generation time compared to conventional binary population synthesis codes. However, there are important potential biases that one must take into account, especially those introduced by the forward-simulator and data summarizer. The posterior probability of the SBI can be explicitly written as $p(\boldsymbol{\Lambda} \vert \boldsymbol{s}(\boldsymbol{D}),\, \boldsymbol{M}_\rm{Sim})$, where $\boldsymbol{s}(\boldsymbol{D})$ is the summary of the data $\boldsymbol{D}$, and $\boldsymbol{M}_\rm{Sim}$ represents the forward simulator model. The accuracy of the posteriors is also conditioned on that of the forward simulator (described in \S~\ref{sec:method:pop-sim},~\ref{sec:method:data-sim}). An SBI trained on simulation data that does not accurately model observational data (both signal and noise characteristics) will produce inaccurate posteriors. 
Furthermore, the choice of summary statistics that we define in \S~\ref{sec:method:data-summary} can potentially skew the information from the data that is passed to the neural network. As the posteriors of the SBI are conditioned on the summary, and not the data itself, a biased/ill-informed summary can therefore bias posteriors. 

In future iterations, we will explore the robustness of the data summary to different simulation models \citep{RobustSummaries_ModelMispec_2023} and test the SBI for model mis-specification \citep{AnauMontel:2024flo}. Mismodeling can arise, for instance, from incorrect/inadequate parametric representation of the population parameters (eg., mass and separation distribution), the absence of simulating overlapping signals from other sources (like MBHBs, EMRIs, stBHBs), and from simplified assumptions on the noise (stationary, Gaussianity), which may not hold in real data.

\section{Where do constraints come from?}
\label{sec:discussion:WhereInfo}
An important question regarding Galactic binaries is whether the information signature for population inference resides in the resolvable or the unresolvable sources. To shed light on this, we train and test two separate SBIs with identical network architectures on the two classes of sources. $\SBIres$ ($\SBIunres$) is trained to estimate the posteriors of the population from LISA's channel-A data containing only instrumental noise and (un)resolvable DWDs. A DWD is deemed resolvable if its single detector optimal match-filter SNR $\rho_{\text{opt}}$ is greater than 7. $\rho_{\text{opt}}$ is computed assuming a fiducial noise PSD with $\A = \O = 1$, augmented by the stochastic DWD background as parametrized in Ref.~\cite{LDC2022}. The results presented below and the conclusions drawn do not qualitatively depend on this SNR threshold.

\begin{figure*}
    \centering
    \includegraphics[width=\linewidth]{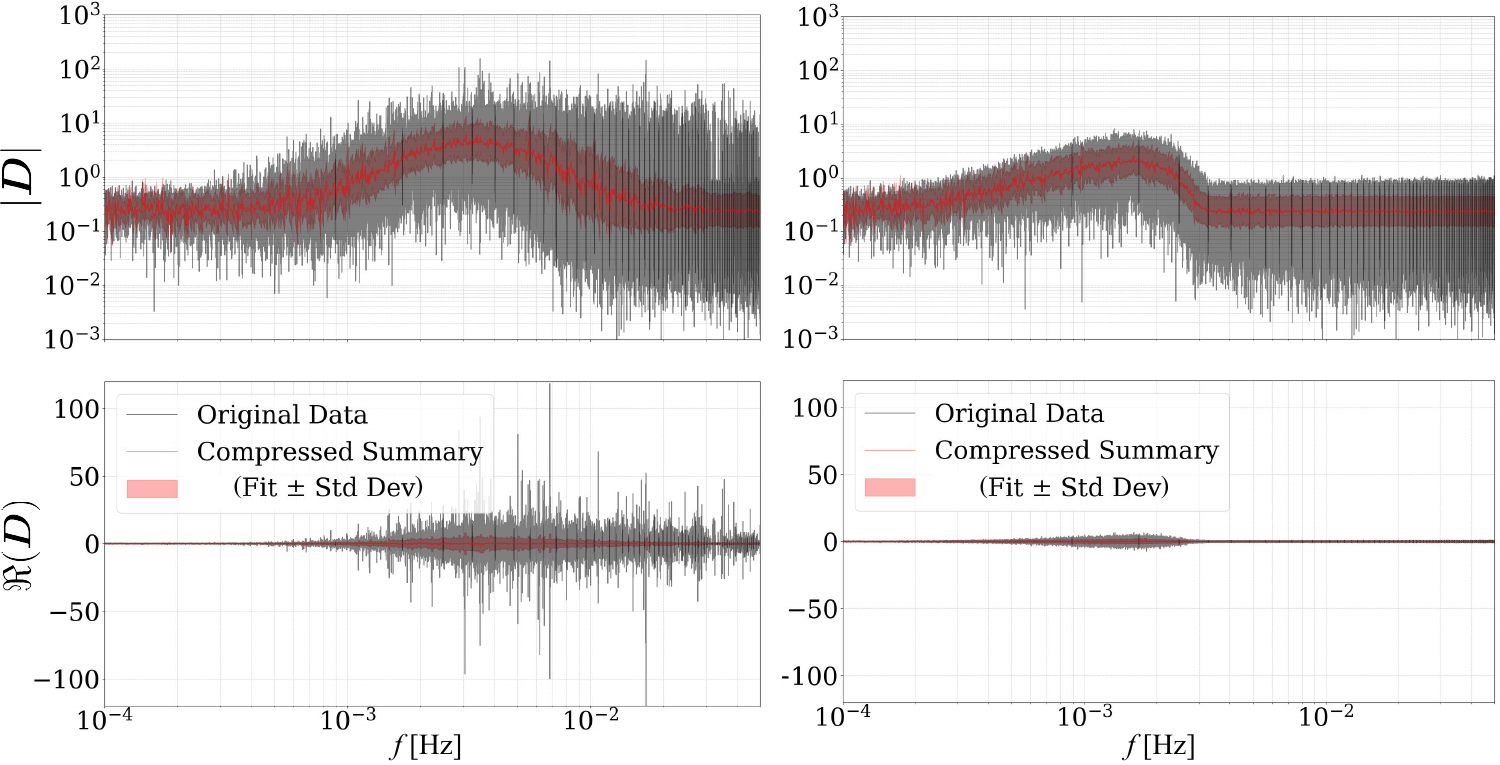}
    \caption{Data including only resolvable (left) and only unresolvable (right) sources. Parameters are fixed to their fiducial value.
    }
    \label{fig:Data_ResolveUnResolve}
\end{figure*}

In Fig.~\ref{fig:Data_ResolveUnResolve}, we present the data provided to the two SBIs, which share the same instrumental noise and Galactic population as those used to generate Fig.~\ref{fig:posterior}, filtered into their respective disjoint subpopulations of resolvable and unresolvable sources. In comparison to the unresolvable case, the resolvable sources produce raw, uncompressed (black) data with large variations. This is also reflected by the larger standard deviations in the compressed (red) data. Although resolvable sources are three orders of magnitude fewer than unresolvable ones, the bump observed in Fig.~\ref{fig:data} is primarily contributed by the resolvable sources as evident from Fig.~\ref {fig:Data_ResolveUnResolve}.

\begin{figure*}
    \centering
    \includegraphics[width=\linewidth]{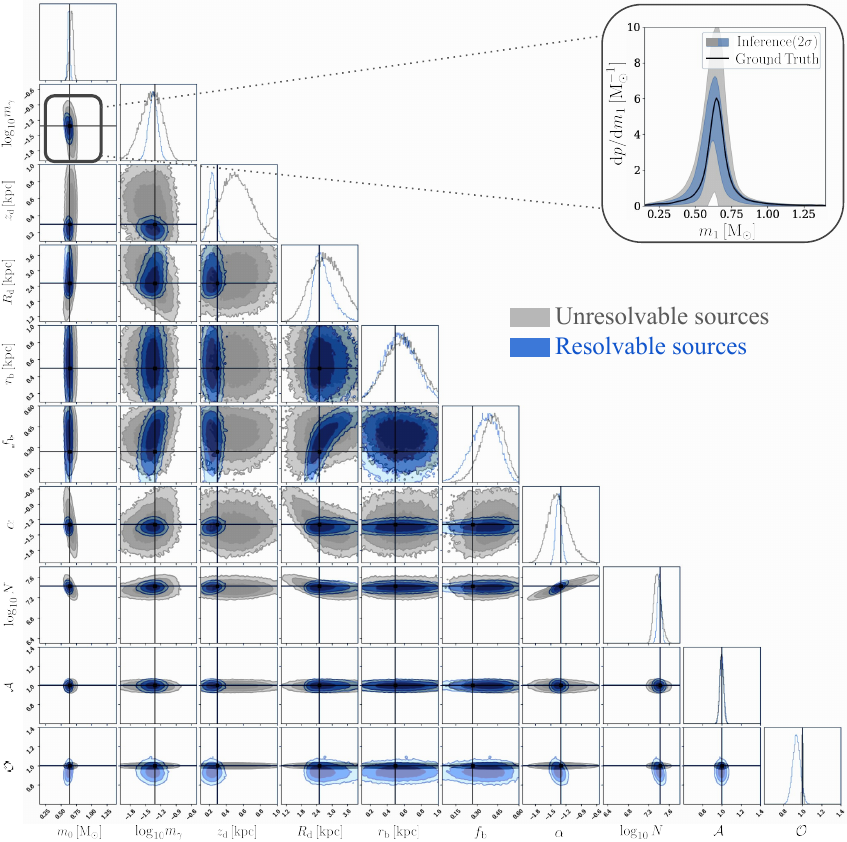}
    \caption{Posterior for the unresolvable (gray) and resolvable (blue) sources of the astrophysical population analyzed in Fig.~\ref{fig:posterior}. Black lines indicate the ground truth. The inset shows the 2$\sigma$ contours of the two populations in their respective colors. }
    \label{fig:corner_resolve}
\end{figure*}

Passing this data to both $\SBIres$ and $\SBIunres$, Fig.~\ref{fig:corner_resolve} compares their A channel posteriors. All parameters, barring high-frequency noise component $\O$, are better constrained by the resolvable subpopulation. Moreover, in comparison to $\SBIunres$, the posteriors from $\SBIres$ are more alike those from the total population in Fig.~\ref{fig:posterior}. The $\SBIunres$ posteriors of the mass distribution parameters, Galactic scale radius, and separation index are far more diffuse in comparison to those of $\SBIres$. These results, along with comparable $\logN, \, \A$ posteriors, suggest that the resolvable sources form a sufficiently representative subsample of the total population to produce informative posteriors. The improved posteriors of noise parameter $\O$ in $\SBIunres$ as compared to $\SBIres$ is likely due to a more accurate estimation of the noise characteristic at higher frequencies. This likely stems from the fact that in the case of the former, there are no sharp fluctuations caused by resolvable DWDs, especially at higher frequencies, which might otherwise be misattributed to noise. 

Finally we also note that Fig.~\ref{fig:corner_resolve} shows a slight negative correlation between the posteriors of $\m0$ and $\alpha$, a negative correlation with $\logN$. These correlations are more prominent at lower SNRs and hence not seen in the joint population of Fig.~\ref{fig:posterior}. DWD populations from different mass distributions occupy different frequency regions (c.f. the frequency histogram in Fig.~\ref{fig:pop}). Populations with a higher median mass (i.e., larger $\m0$) have lower GW frequencies. Due to the lower PSD of LISA at these frequencies (c.f. the PSD in Fig.~\ref{fig:pop}), this results in lower SNR. Thus, the correlation arises because of the similar effect that increasing $\logN$ and decreasing $\m0$ has on the SNR of the sources.

It is important to note that the conclusions drawn here are conditioned on the specific summary statistics used; a different summary that captures additional features from the unresolved sources could alter this interpretation. Our compression method summarizes individually resolvable source peaks and the unresolvable power spectrum within a frequency bin by a linear fit, the standard deviation, and outlier power. Summaries with smaller frequency bins (greater resolution) and/or non-linear functions might present greater fidelity to the detector data. Inference from high-fidelity summaries, assuming a sufficiently expressive SBI and adequate training data, would be more accurate, potentially with better coverage before calibration. This is especially true in the analysis of GBs, given that a large fraction of the information appears to come from the resolvable peaks. However, with increasingly larger summaries, the number of network parameters increases rapidly, requiring proportionally large GPU VRAM. The degree of data compression is thus a tradeoff between fidelity towards input data and computational limitations.

\section{Conclusion}
\label{sec:conclusion}
In this work, we have introduced an amortized SBI framework to extract the population properties of Galactic DWDs from LISA frequency (or time) series data, without the traditional, computationally expensive DWD source-by-source extraction of the global-fit. 

We integrate a fast, GPU-accelerated forward simulator of the DWD population and its associated GW signals to generate several hundred thousand training examples. A conditional normalizing flow is then trained on summary statistics that capture both the collective signature of the unresolved sub-population and the features of the individually resolvable sources.
The summarizer takes as input the absolute amplitude and real parts of the frequency spectrum, compressing the large number ($\mathcal{O}(10^7)$) of frequency values into a much smaller set of summary statistics ($\mathcal{O}(6\times10^3)$). This enables rapid amortized posterior estimation of key population parameters, including the primary mass distribution, the Milky Way's disk and bulge scale lengths, the bulge to disk fraction, the DWD separation distribution power-law index, and the total number of DWDs in the Galaxy.

Importantly, our analysis utilizes information from both the relatively few ($\mathcal{O}(10^4)$) resolved sources and the vastly more numerous ($\mathcal{O}(10^7)$) unresolved foreground sources. By using the full data, we naturally mitigate the selection biases that typically affect hierarchical inference based on finite detection catalogs, as in standard global-fit methods that focus only (at first) on resolvable sources. In the context of Galactic binary population inference, our proof of concept promises to complement the global-fit, at a fraction of the computation time. Moreover, once trained and applied to real LISA data, our SBI framework can provide instantaneous population-level constraints as new observations are collected, enabling continuously updated, real-time population inference throughout the mission.

Our results show that the calibrated SBI approach delivers rapid posterior estimates with (approximately) unbiased coverage, successfully recovering the injected population parameters with percent-level precision for the primary mass distribution, separation index, and total number of sources, and with tens-of-percent precision for the Galactic disk scales.  Crucially, by including both amplitude and phase information in the summary statistics, we can break degeneracies that would otherwise leave the spatial-scale parameters prior-dominated. 

We carried out an investigation of whether the information relevant for Galactic binary population inference resides primarily in resolvable or unresolvable sources. By training two separate SBI models on disjoint subpopulations --one containing only resolvable binaries, the other only unresolvable ones-- we probed the information content of each class. Despite being outnumbered by unresolvable sources by three orders of magnitude, the resolvable binaries yielded significantly tighter constraints on the population parameters. Their posteriors closely match those inferred from the full population, indicating that resolvable sources form an informative and representative subsample for constraining key population parameters. Conversely, the unresolvable subpopulation yielded broader posteriors, though it contributed to more accurate estimation of instrumental noise characteristics, which likely stems from the absence of sharp spectral features associated with high-SNR sources. Overall, our findings highlight the dominant role of resolvable binaries in constraining Galactic population properties. To our knowledge, this is the first systematic, simulation-based analysis of the population signal content across resolvability classes, offering a novel perspective on where the population-level information truly resides in LISA data.

An important caveat of parametric reconstruction techniques—including SBI as well as tractable likelihood methods—is that the inferred posteriors are only as accurate as the forward model used to simulate the data; that is, the posterior probability is predicated on the assumption that the forward model is correct. It is therefore essential to perform inference using a variety of forward simulators to assess the robustness of the results. To this end, we emphasize that our framework is inherently flexible and can readily incorporate additional source classes (e.g., MBHBs, stBHBs, EMRIs) as well as un-modeled sources different noise features, including non-Gaussianity and time-dependent artifacts, such as data gaps and glitches. Modeling these features is necessary for both simulation-based and global fit methods.

Future work may explore the use of SBI performed sequentially in time as LISA observations accumulate, providing real-time updates on source and noise parameter inference. Finally, in future iterations of the method, we may explore different or refined summary statistics, which can be learned via neural autoencoders to improve sensitivity to subtle population features, or different
SBI methods such as truncated marginal neural ratio estimation (see e.g. Refs.~\cite{Perigrine_2023,Alvey:2023npw,Cole:2025sqo} for  applications to GW data analysis).

\textbf{Data Availability:  }
The code used to reproduce the results presented in this paper and to construct the simulation-based inference pipeline will be made publicly available shortly.

\begin{acknowledgements}
We thank Chantal Pitte and Neil Cornish for their useful suggestions and comments, and Valeriya Korol for providing the code for the population synthesis. We acknowledge support from the European Union’s H2020 ERC Consolidator Grant ``GRavity from Astrophysical to Microscopic Scales'' (Grant No. GRAMS-815673, to E.B. and R.S.), the European Union’s Horizon ERC Synergy Grant ``Making Sense of the Unexpected in the Gravitational-Wave Sky'' (Grant No. GWSky-101167314, to E.B. and R.S.), the PRIN 2022 grant ``GUVIRP - Gravity tests in the UltraViolet and InfraRed with Pulsar timing'' (to E.B.), and the EU Horizon 2020 Research and Innovation Programme under the Marie Sklodowska-Curie Grant Agreement No. 101007855 (to E.B.). 
R.T. acknowledges co-funding from Next Generation EU, in the context of the National Recovery and Resilience Plan, Investment PE1 Project FAIR ``Future Artificial Intelligence Research''. This resource was co-financed by the Next Generation EU [DM 1555 del 11.10.22]. R.T. is partially supported by the Fondazione ICSC, Spoke 3 ``Astrophysics and Cosmos Observations'', Piano Nazionale di Ripresa e Resilienza Project ID CN00000013 ``Italian Research Center on High-Performance Computing, Big Data and Quantum Computing'' funded by MUR Missione 4 Componente 2 Investimento 1.4: Potenziamento strutture di ricerca e creazione di ``campioni nazionali di R\&S (M4C2-19)'' - Next Generation EU (NGEU). 
This work has been supported by the Agenzia Spaziale Italiana (ASI), Project n. 2024-36-HH.0, ``Attività per la fase B2/C della missione LISA''. Computational work has been made possible through SISSA-CINECA and CINECA-INFN agreements providing access to resources on LEONARDO at CINECA.MB.
N.K. acknowledges support from the CNES for the exploration of LISA science. N.K. gratefully acknowledges support from the CNRS/IN2P3 Computing Center (Lyon - France) for providing computing and data-processing resources needed for this work. 
\end{acknowledgements}
\vskip 0.5cm
\appendix 
\label{sec:appendix}

\section{Primary mass fit}
\begin{figure}
    \centering
    \includegraphics[width=.7\linewidth]{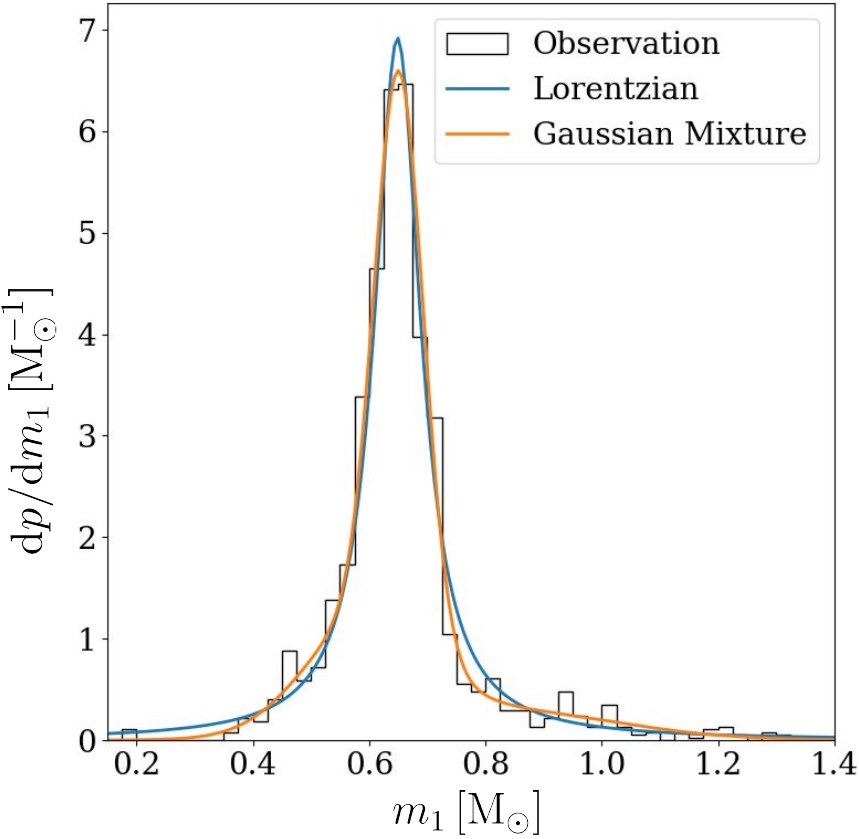}
    \caption{Comparison of the eight-parameter Gaussian mixture fit \cite{Kepler+2015} and our two-parameter Lorentzian fit of the observed white dwarf mass distribution.}
    \label{fig:mass-fit}
\end{figure}

Instead of the Gaussian mixture model in \cite{Maoz+2018} (requiring 8 free parameters), we model the primary mass $m_1$ of the DWD as a truncated Lorentzian (Cauchy) probability distribution (two free parameters) given by

\be
\frac{dp}{dm_1} \propto
\begin{cases}
 \left[\pi \mgamma \left(1 + \left( \frac{x - \m0}{\mgamma} \right)^2 \right) \right]^{-1}  \\\text{ for }0.15 \Msun \leq m_1 \leq  1.45 \Msun \\\\
0 \,\,\, \text{otherwise},
\end{cases}
\label{eq:mass-fit}
\ee 
where $\m0$ is the mass corresponding to the peak of the distribution and $\mgamma$ corresponds to the half-width half-maximum value. We sample $\m0$ from a uniform distribution. For better representation of both narrow and flat mass distributions, we sample $\mgamma$ from a log-uniform prior ($\logmgamma$ sampled uniformly). In Fig.~\ref{fig:mass-fit}, we compare our fit with that of the observation data and the Gaussian mixture model from \cite{Maoz+2018}, showing a reasonable overlap of all three distributions. 

\label{sec:appendix:Mass_fit}

\begin{figure*}
    \centering
    \includegraphics[width=.7\linewidth]{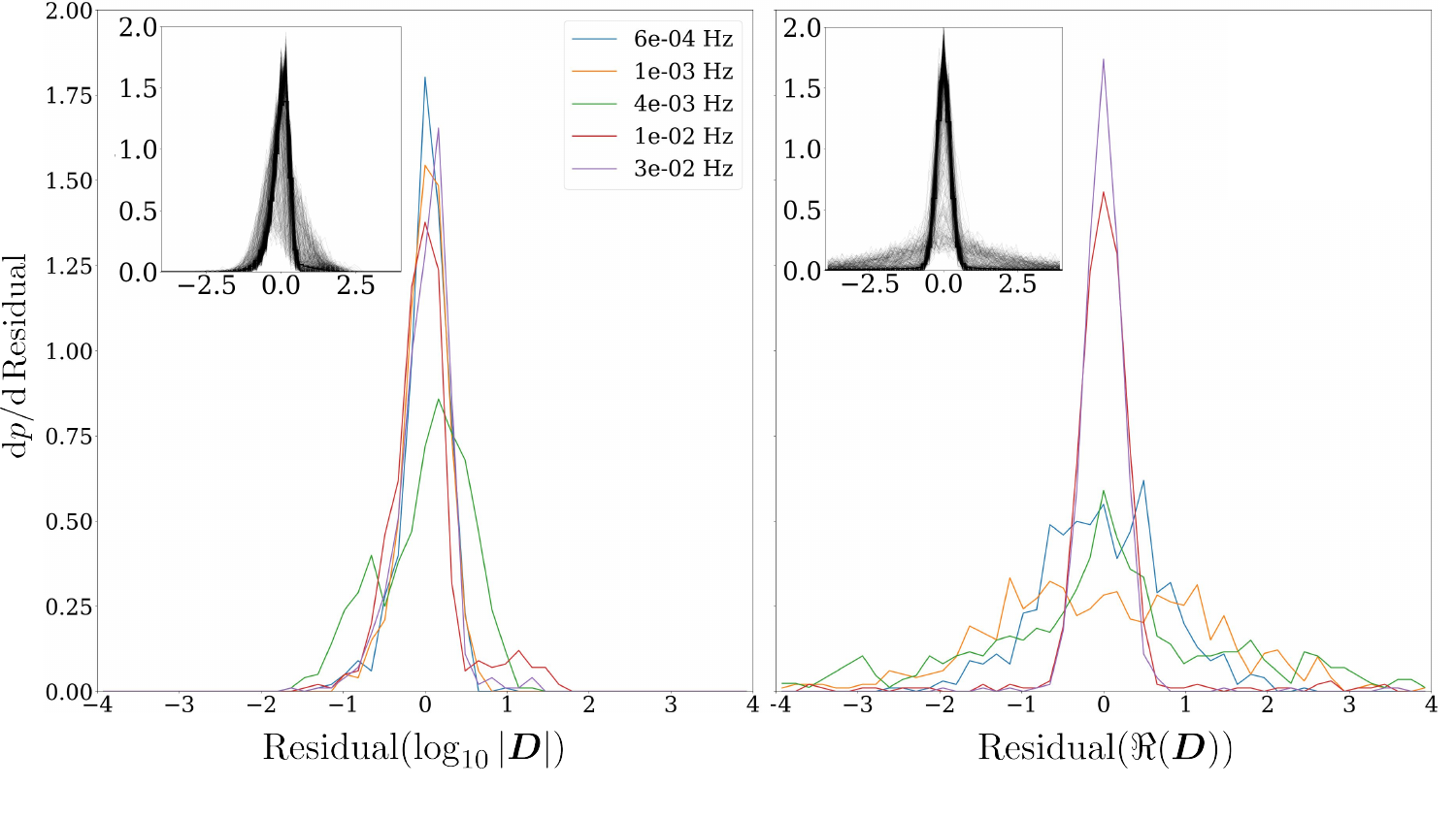}
    \caption{The distribution of the residual across the training dataset for a few example frequency bins (colored lines, see legend) and for all bins combined (inset) for the summaries of the logarithm of the absolute $\boldsymbol{s}(\log_{10}\vert \boldsymbol{D}\vert)$ (left) and real $\boldsymbol{s}(\Re(\boldsymbol{D)})$ (right) data.}
    \label{fig:residual}
\end{figure*}

\begin{table*}[ht]
\centering
\begin{tabular}{lccc}
\hline
\textbf{Layer} & \textbf{Input Shape} & \textbf{Output Shape} & \textbf{Details} \\
\hline
Input Reshape & $(B, 8192)$ & $(B, 8, 1024)$ & Reshape to 8-channel input \\
Conv1d + BN + ReLU & $(B, 8, 1024)$ & $(B, 32, 512)$ & Kernel: 8, Stride: 2, Padding: 3 \\
Conv1d + BN + ReLU & $(B, 32, 512)$ & $(B, 64, 256)$ & Kernel: 8, Stride: 2, Padding: 3 \\
Conv1d + BN + ReLU & $(B, 64, 256)$ & $(B, 128, 128)$ & Kernel: 8, Stride: 2, Padding: 3 \\
Conv1d + BN + ReLU & $(B, 128, 128)$ & $(B, 256, 64)$ & Kernel: 8, Stride: 2, Padding: 3 \\
Reshape (Flatten) + Dropout & - & $(B, 16384)$ & Dropout probability $=0.2$ \\
Fully Connected + ReLU + Dropout & $(B, 16384)$ & $(B, 256)$ & Dropout probability $=0.2$ \\
Fully Connected & $(B, 256)$ & $(B, 256)$ & Context vector passed to normalizing flow\\
\hline
\end{tabular}
\caption{Architecture of \texttt{CNN\_Encoder}, a convolutional neural network encoder used to construct a latent representation of the frequency-domain input. B denotes the batch size; Conv1d refers to a one-dimensional convolutional layer; BN denotes batch normalization; ReLU is the rectified linear unit activation function; and fully connected refers to a multilayer perception layer.}
\label{tab:CNN}
\end{table*}

\begin{table*}
\centering
\begin{tabular}{lccc}
\hline
\textbf{Parameter} & \textbf{Value} & \textbf{Description} \\
\hline
Context encoder       & -          & \texttt{CNN\_Encoder}\\
Number of Transforms   & 16        & Number of MAF layers \\
Hidden Features        & 512       & Width of each autoregressive NN \\
Number of Blocks       & 4         & Layers per NN inside each transform \\
Dropout probability    & 0.05      & Prevents overfitting \\
\hline
\end{tabular}
\caption{Architecture of \texttt{NF}, the Masked Autoregressive Flow (MAF) density estimator.}
\label{tab:NF}
\end{table*}

\begin{figure*}
    \centering
    \includegraphics[width=\linewidth]{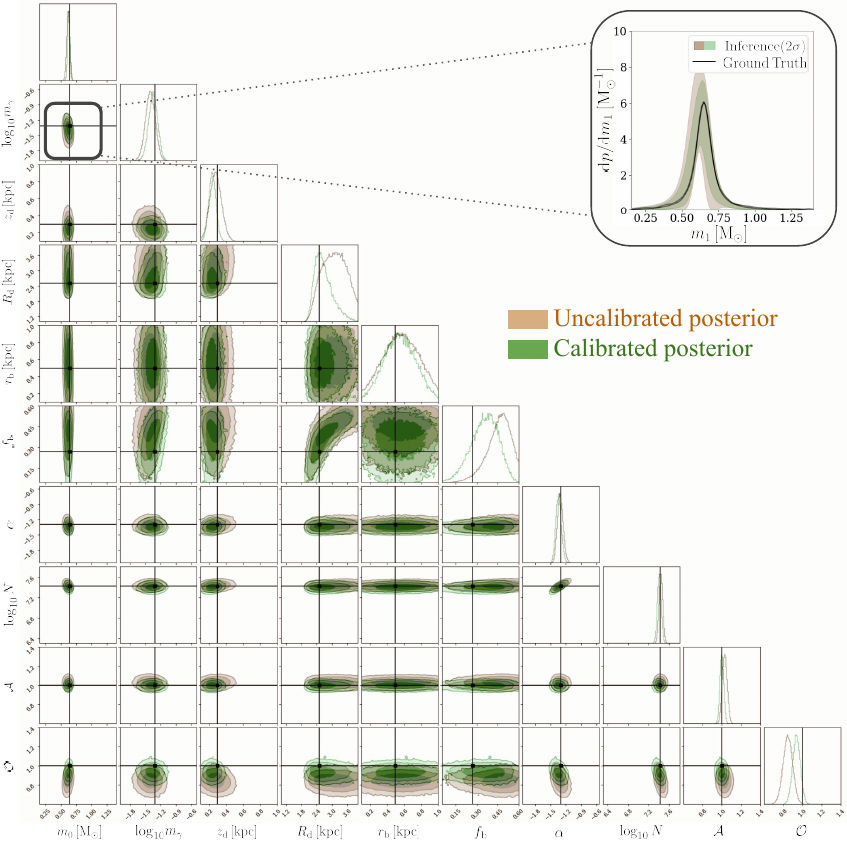}
    \caption{Comparison of the uncalibrated and calibrated posteriors for the same population realization as that of Fig.~\ref{fig:posterior}.}
    \label{fig:posterior_uncalib_calib}
\end{figure*}

\section{Residual distribution}
\label{sec:appendix:Residual_distribution}
In Fig.~\ref{fig:residual}, we show the distribution of the residual of the data and the linear fit at different (and all) frequency bins. We summarize the bulk of the generally symmetric, zero-centered distribution by its standard deviation and model the outlier tails by their $L_2$ norm.

\section{Network architecture}
\label{sec:appendix:nn}
In this section, we describe the architecture of the SBI's conditional normalizing flow network. The normalizing flow \texttt{NF} is conditioned on the context vector, a latent representation of the data extracted by a convolutional neural network \texttt{CNN\_Encoder}.

\section{Calibrated posteriors}
\label{sec:appendix:CalibrationComparison}
Following the temperature scaling procedure outlined in $\S$~\ref{sec:method:calibration}, we resample the uncalibrated posterior samples to obtain the calibrated distribution. Each uncalibrated sample $i$, with value $\boldsymbol{\Lambda}_i$ and associated probability $p_{\rm SBI,\,i}$, is assigned a weight $w_i = p_{\rm SBI,\,i} \,^{1 - 1/T^*}$. In Fig.~\ref{fig:posterior_uncalib_calib}, we compare the two distributions. The calibrated posteriors are generally narrower in comparison to the uncalibrated due to the underconfident coverage shown in the PP plot of Fig.~\ref{fig:coverage}.

\clearpage
\bibliographystyle{utphys}
\bibliography{refs}

@article{Karchev:2022xyn,
    author = "Karchev, Konstantin and Trotta, Roberto and Weniger, Christoph",
    title = "{SICRET: Supernova Ia Cosmology with truncated marginal neural Ratio EsTimation}",
    eprint = "2209.06733",
    archivePrefix = "arXiv",
    primaryClass = "astro-ph.CO",
    doi = "10.1093/mnras/stac3785",
    journal = "Mon. Not. Roy. Astron. Soc.",
    volume = "520",
    number = "1",
    pages = "1056--1072",
    year = "2023"
}

@article{Korol+2022,
    author = {Korol, Valeriya and Hallakoun, Na’ama and Toonen, Silvia and Karnesis, Nikolaos},
    title = {Observationally driven Galactic double white dwarf population for LISA},
    journal = {Monthly Notices of the Royal Astronomical Society},
    volume = {511},
    number = {4},
    pages = {5936-5947},
    year = {2022},
    month = {02},
    issn = {0035-8711},
    doi = {10.1093/mnras/stac415}
}

@article{Deng:2025wgk,
    author = "Deng, Senwen and Babak, Stanislav and Le Jeune, Maude and Marsat, Sylvain and Plagnol, \'Eric and Sartirana, Andrea",
    title = "{Modular global-fit pipeline for LISA data analysis}",
    eprint = "2501.10277",
    archivePrefix = "arXiv",
    primaryClass = "gr-qc",
    month = "1",
    year = "2025"
}

@article{Littenberg:2023xpl,
    author = "Littenberg, Tyson B. and Cornish, Neil J.",
    title = "{Prototype global analysis of LISA data with multiple source types}",
    eprint = "2301.03673",
    archivePrefix = "arXiv",
    primaryClass = "gr-qc",
    doi = "10.1103/PhysRevD.107.063004",
    journal = "Phys. Rev. D",
    volume = "107",
    number = "6",
    pages = "063004",
    year = "2023"
}

@article{Strub:2024kbe,
    author = {Strub, Stefan H. and Ferraioli, Luigi and Schmelzbach, C\'edric and St\"ahler, Simon C. and Giardini, Domenico},
    title = "{Global analysis of LISA data with Galactic binaries and massive black hole binaries}",
    eprint = "2403.15318",
    archivePrefix = "arXiv",
    primaryClass = "gr-qc",
    doi = "10.1103/PhysRevD.110.024005",
    journal = "Phys. Rev. D",
    volume = "110",
    number = "2",
    pages = "024005",
    year = "2024"
}

@article{Katz:2024oqg,
    author = "Katz, Michael L. and Karnesis, Nikolaos and Korsakova, Natalia and Gair, Jonathan R. and Stergioulas, Nikolaos",
    title = "{Efficient GPU-accelerated multisource global fit pipeline for LISA data analysis}",
    eprint = "2405.04690",
    archivePrefix = "arXiv",
    primaryClass = "gr-qc",
    doi = "10.1103/PhysRevD.111.024060",
    journal = "Phys. Rev. D",
    volume = "111",
    number = "2",
    pages = "024060",
    year = "2025"
}

@article{Adams:2012qw,
    author = "Adams, Matthew R. and Cornish, Neil J. and Littenberg, Tyson B.",
    title = "{Astrophysical Model Selection in Gravitational Wave Astronomy}",
    eprint = "1209.6286",
    archivePrefix = "arXiv",
    primaryClass = "gr-qc",
    doi = "10.1103/PhysRevD.86.124032",
    journal = "Phys. Rev. D",
    volume = "86",
    pages = "124032",
    year = "2012"
}

@ARTICLE{Cornish_Littenberg_2007,
       author = {{Cornish}, Neil J. and {Littenberg}, Tyson B.},
        title = "{Tests of Bayesian model selection techniques for gravitational wave astronomy}",
      journal = {\prd},
         year = 2007,
        month = oct,
       volume = {76},
       number = {8},
          eid = {083006},
        pages = {083006},
          doi = {10.1103/PhysRevD.76.083006},
archivePrefix = {arXiv},
       eprint = {0704.1808},
 primaryClass = {gr-qc},
       adsurl = {https://ui.adsabs.harvard.edu/abs/2007PhRvD..76h3006C}
}

@article{Spadaro:2023muy,
    author = "Spadaro, Alice and Buscicchio, Riccardo and Vetrugno, Daniele and Klein, Antoine and Vitale, Stefano and Dolesi, Rita and Weber, William Joseph and Colpi, Monica",
    title = "{Glitch systematics on the observation of massive black-hole binaries with LISA}",
    eprint = "2306.03923",
    archivePrefix = "arXiv",
    primaryClass = "gr-qc",
    doi = "10.1103/PhysRevD.108.123029",
    journal = "Phys. Rev. D",
    volume = "108",
    number = "12",
    pages = "123029",
    year = "2023"
}

@article{Burke:2025bun,
    author = "Burke, Ollie and Marsat, Sylvain and Gair, Jonathan R. and Katz, Michael L.",
    title = "{Mind the gap: addressing data gaps and assessing noise mismodeling in LISA}",
    eprint = "2502.17426",
    archivePrefix = "arXiv",
    primaryClass = "gr-qc",
    month = "2",
    year = "2025"
}

@article{LISA:2022yao,
    author = "Seoane, Pau Amaro and others",
    collaboration = "LISA",
    title = "{Astrophysics with the Laser Interferometer Space Antenna}",
    eprint = "2203.06016",
    archivePrefix = "arXiv",
    primaryClass = "gr-qc",
    doi = "10.1007/s41114-022-00041-y",
    journal = "Living Rev. Rel.",
    volume = "26",
    number = "1",
    pages = "2",
    year = "2023"
}

@article{Korol_Rossi_Barausse_2018,
    author = "Korol, Valeriya and Rossi, Elena M. and Barausse, Enrico",
    title = "{A multimessenger study of the Milky Way\textquoteright{}s stellar disc and bulge with LISA, Gaia, and LSST}",
    eprint = "1806.03306",
    archivePrefix = "arXiv",
    primaryClass = "astro-ph.GA",
    doi = "10.1093/mnras/sty3440",
    journal = "Mon. Not. Roy. Astron. Soc.",
    volume = "483",
    number = "4",
    pages = "5518--5533",
    year = "2019"
}

@article{Lackeos:2023eub,
    author = "Lackeos, Kristen and Littenberg, Tyson B. and Cornish, Neil J. and Thorpe, James I.",
    title = "{The LISA Data Challenge Radler analysis and time-dependent ultra-compact binary catalogues}",
    eprint = "2308.12827",
    archivePrefix = "arXiv",
    primaryClass = "gr-qc",
    doi = "10.1051/0004-6361/202347222",
    journal = "Astron. Astrophys.",
    volume = "678",
    pages = "A123",
    year = "2023"
}

@article{Johnson:2025oyu,
    author = "Johnson, Aaron D. and Roulet, Javier and Chatziioannou, Katerina and Vallisneri, Michele and Trejo, Chris G. and Gersbach, Kyle A.",
    title = "{PETRA: From the global fit for LISA's Galactic binaries to a catalog of sources}",
    eprint = "2502.14818",
    archivePrefix = "arXiv",
    primaryClass = "gr-qc",
    month = "2",
    year = "2025"
}

@article{Littenberg:2020bxy,
    author = "Littenberg, Tyson and Cornish, Neil and Lackeos, Kristen and Robson, Travis",
    title = "{Global Analysis of the Gravitational Wave Signal from Galactic Binaries}",
    eprint = "2004.08464",
    archivePrefix = "arXiv",
    primaryClass = "gr-qc",
    doi = "10.1103/PhysRevD.101.123021",
    journal = "Phys. Rev. D",
    volume = "101",
    number = "12",
    pages = "123021",
    year = "2020"
}

@article{Lamberts:2019nyk,
    author = "Lamberts, Astrid and Blunt, Sarah and Littenberg, Tyson B. and Garrison-Kimmel, Shea and Kupfer, Thomas and Sanderson, Robyn E.",
    title = "{Predicting the LISA white dwarf binary population in the Milky Way with cosmological simulations}",
    eprint = "1907.00014",
    archivePrefix = "arXiv",
    primaryClass = "astro-ph.HE",
    doi = "10.1093/mnras/stz2834",
    journal = "Mon. Not. Roy. Astron. Soc.",
    volume = "490",
    number = "4",
    pages = "5888--5903",
    year = "2019"
}

@article{Korol:2017qcx,
    author = "Korol, Valeriya and Rossi, Elena M. and Groot, Paul J. and Nelemans, Gijs and Toonen, Silvia and Brown, Anthony G. A.",
    title = "{Prospects for detection of detached double white dwarf binaries with Gaia, LSST and LISA}",
    eprint = "1703.02555",
    archivePrefix = "arXiv",
    primaryClass = "astro-ph.HE",
    doi = "10.1093/mnras/stx1285",
    journal = "Mon. Not. Roy. Astron. Soc.",
    volume = "470",
    number = "2",
    pages = "1894--1910",
    year = "2017"
}

@ARTICLE{Maoz+2018,
       author = {{Maoz}, Dan and {Hallakoun}, Na'ama and {Badenes}, Carles},
        title = "{The separation distribution and merger rate of double white dwarfs: improved constraints}",
      journal = {\mnras},
         year = 2018,
        month = may,
       volume = {476},
       number = {2},
        pages = {2584-2590},
          doi = {10.1093/mnras/sty339},
archivePrefix = {arXiv},
       eprint = {1801.04275},
 primaryClass = {astro-ph.SR},
       adsurl = {https://ui.adsabs.harvard.edu/abs/2018MNRAS.476.2584M}
}

@ARTICLE{Kepler+2015,
       author = {{Kepler}, S.~O. and {Pelisoli}, I. and {Koester}, D. and {Ourique}, G. and {Kleinman}, S.~J. and {Romero}, A.~D. and {Nitta}, A. and {Eisenstein}, D.~J. and {Costa}, J.~E.~S. and {K{\"u}lebi}, B. and {Jordan}, S. and {Dufour}, P. and {Giommi}, Paolo and {Rebassa-Mansergas}, Alberto},
        title = "{New white dwarf stars in the Sloan Digital Sky Survey Data Release 10}",
      journal = {\mnras},
         year = 2015,
        month = feb,
       volume = {446},
       number = {4},
        pages = {4078-4087},
          doi = {10.1093/mnras/stu2388},
archivePrefix = {arXiv},
       eprint = {1411.4149},
 primaryClass = {astro-ph.SR},
       adsurl = {https://ui.adsabs.harvard.edu/abs/2015MNRAS.446.4078K}
}

@ARTICLE{Moe_DiStefano_2017,
       author = {{Moe}, Maxwell and {Di Stefano}, Rosanne},
        title = "{Mind Your Ps and Qs: The Interrelation between Period (P) and Mass-ratio (Q) Distributions of Binary Stars}",
      journal = {\apjs},
         year = 2017,
        month = jun,
       volume = {230},
       number = {2},
          eid = {15},
        pages = {15},
          doi = {10.3847/1538-4365/aa6fb6},
archivePrefix = {arXiv},
       eprint = {1606.05347},
 primaryClass = {astro-ph.SR},
       adsurl = {https://ui.adsabs.harvard.edu/abs/2017ApJS..230...15M}
}

@ARTICLE{Duchene_Kraus_2013,
       author = {{Duch{\^e}ne}, Gaspard and {Kraus}, Adam},
        title = "{Stellar Multiplicity}",
      journal = {\araa},
         year = 2013,
        month = aug,
       volume = {51},
       number = {1},
        pages = {269-310},
          doi = {10.1146/annurev-astro-081710-102602},
archivePrefix = {arXiv},
       eprint = {1303.3028},
 primaryClass = {astro-ph.SR},
       adsurl = {https://ui.adsabs.harvard.edu/abs/2013ARA&A..51..269D}
}

@software{nflows,
  author       = {Conor Durkan and
                  Artur Bekasov and
                  Iain Murray and
                  George Papamakarios},
  title        = {{nflows}: normalizing flows in {PyTorch}},
  month        = nov,
  year         = 2020,
  publisher    = {Zenodo},
  version      = {v0.14},
  doi          = {10.5281/zenodo.4296287},
  url          = {https://doi.org/10.5281/zenodo.4296287}
}

@ARTICLE{CE_DWDseperation_Korol+2022,
       author = {{Korol}, Valeriya and {Belokurov}, Vasily and {Toonen}, Silvia},
        title = "{A gap in the double white dwarf separation distribution caused by the common-envelope evolution: astrometric evidence from Gaia}",
      journal = {\mnras},
         year = 2022,
        month = sep,
       volume = {515},
       number = {1},
        pages = {1228-1246},
          doi = {10.1093/mnras/stac1686},
archivePrefix = {arXiv},
       eprint = {2203.03659},
 primaryClass = {astro-ph.SR},
       adsurl = {https://ui.adsabs.harvard.edu/abs/2022MNRAS.515.1228K}
}

@ARTICLE{Delfavero_Breivik+2025,
       author = {{Delfavero}, Vera and {Breivik}, Katelyn and {Thiele}, Sarah and {O'Shaughnessy}, Richard and {Baker}, John G.},
        title = "{Recovering Injected Astrophysics from the LISA Double White Dwarf Binaries}",
      journal = {\apj},
         year = 2025,
        month = mar,
       volume = {981},
       number = {1},
          eid = {66},
        pages = {66},
          doi = {10.3847/1538-4357/ada9e2},
archivePrefix = {arXiv},
       eprint = {2409.15230},
 primaryClass = {gr-qc},
       adsurl = {https://ui.adsabs.harvard.edu/abs/2025ApJ...981...66D}
}

@ARTICLE{TMNRE_Miller+2021,
       author = {{Miller}, Benjamin and {Cole}, Alex and {Forr{\'e}}, Patrick and {Louppe}, Gilles and {Weniger}, Christoph},
        title = "{Truncated Marginal Neural Ratio Estimation}",
      journal = {Advances in Neural Information Processing Systems},
         year = 2021,
        month = oct,
       volume = {34},
        pages = {129},
          doi = {10.48550/arXiv.2107.01214},
archivePrefix = {arXiv},
       eprint = {2107.01214},
 primaryClass = {stat.ML},
       adsurl = {https://ui.adsabs.harvard.edu/abs/2021ANIPS..34..129M}
}

@ARTICLE{Hollands+2018,
       author = {{Hollands}, M.~A. and {Tremblay}, P. -E. and {G{\"a}nsicke}, B.~T. and {Gentile-Fusillo}, N.~P. and {Toonen}, S.},
        title = "{The Gaia 20 pc white dwarf sample}",
      journal = {\mnras},
         year = 2018,
        month = nov,
       volume = {480},
       number = {3},
        pages = {3942-3961},
          doi = {10.1093/mnras/sty2057},
archivePrefix = {arXiv},
       eprint = {1805.12590},
 primaryClass = {astro-ph.SR},
       adsurl = {https://ui.adsabs.harvard.edu/abs/2018MNRAS.480.3942H}
}

@ARTICLE{MWscalebulge_2009,
       author = {{Sofue}, Yoshiaki and {Honma}, Mareki and {Omodaka}, Toshihiro},
        title = "{Unified Rotation Curve of the Galaxy -- Decomposition into de Vaucouleurs Bulge, Disk, Dark Halo, and the 9-kpc Rotation Dip --}",
      journal = {\pasj},
         year = 2009,
        month = feb,
       volume = {61},
        pages = {227},
          doi = {10.1093/pasj/61.2.227},
archivePrefix = {arXiv},
       eprint = {0811.0859},
 primaryClass = {astro-ph},
       adsurl = {https://ui.adsabs.harvard.edu/abs/2009PASJ...61..227S}
}

@ARTICLE{MWscaleheight_2017,
       author = {{Mackereth}, J. Ted and {Bovy}, Jo and {Schiavon}, Ricardo P. and {Zasowski}, Gail and {Cunha}, Katia and {Frinchaboy}, Peter M. and {Garc{\'\i}a Perez}, Ana E. and {Hayden}, Michael R. and {Holtzman}, Jon and {Majewski}, Steven R. and {M{\'e}sz{\'a}ros}, Szabolcs and {Nidever}, David L. and {Pinsonneault}, Marc and {Shetrone}, Matthew D.},
        title = "{The age-metallicity structure of the Milky Way disc using APOGEE}",
      journal = {\mnras},
         year = 2017,
        month = nov,
       volume = {471},
       number = {3},
        pages = {3057-3078},
          doi = {10.1093/mnras/stx1774},
archivePrefix = {arXiv},
       eprint = {1706.00018},
 primaryClass = {astro-ph.GA},
       adsurl = {https://ui.adsabs.harvard.edu/abs/2017MNRAS.471.3057M}
}

@ARTICLE{LSIA_redbook_2024,
       author = {{Colpi}, Monica and {Danzmann}, Karsten and {Hewitson}, Martin and {Holley-Bockelmann}, Kelly and {Jetzer}, Philippe and {Nelemans}, Gijs and {Petiteau}, Antoine and {Shoemaker}, David and {Sopuerta}, Carlos and {Stebbins}, Robin and {Tanvir}, Nial and {Ward}, Henry and {Weber}, William Joseph and {Thorpe}, Ira and {Daurskikh}, Anna and {Deep}, Atul and {Fern{\'a}ndez N{\'u}{\~n}ez}, Ignacio and {Garc{\'\i}a Marirrodriga}, C{\'e}sar and {Gehler}, Martin and {Halain}, Jean-Philippe and {Jennrich}, Oliver and {Lammers}, Uwe and {Larra{\~n}aga}, Jonan and {Lieser}, Maike and {L{\"u}tzgendorf}, Nora and {Martens}, Waldemar and {Mondin}, Linda and {Piris Ni{\~n}o}, Ana and {Amaro-Seoane}, Pau and {Arca Sedda}, Manuel and {Auclair}, Pierre and {Babak}, Stanislav and {Baghi}, Quentin and {Baibhav}, Vishal and {Baker}, Tessa and {Bayle}, Jean-Baptiste and {Berry}, Christopher and {Berti}, Emanuele and {Boileau}, Guillaume and {Bonetti}, Matteo and {Brito}, Richard and {Buscicchio}, Riccardo and {Calcagni}, Gianluca and {Capelo}, Pedro R. and {Caprini}, Chiara and {Caputo}, Andrea and {Castelli}, Eleonora and {Chen}, Hsin-Yu and {Chen}, Xian and {Chua}, Alvin and {Davies}, Gareth and {Derdzinski}, Andrea and {Domcke}, Valerie Fiona and {Doneva}, Daniela and {Dvorkin}, Irna and {Mar{\'\i}a Ezquiaga}, Jose and {Gair}, Jonathan and {Haiman}, Zoltan and {Harry}, Ian and {Hartwig}, Olaf and {Hees}, Aurelien and {Heffernan}, Anna and {Husa}, Sascha and {Izquierdo-Villalba}, David and {Karnesis}, Nikolaos and {Klein}, Antoine and {Korol}, Valeriya and {Korsakova}, Natalia and {Kupfer}, Thomas and {Laghi}, Danny and {Lamberts}, Astrid and {Larson}, Shane and {Le Jeune}, Maude and {Lewicki}, Marek and {Littenberg}, Tyson and {Madge}, Eric and {Mangiagli}, Alberto and {Marsat}, Sylvain and {Vilchez}, Ivan Martin and {Maselli}, Andrea and {Mathews}, Josh and {van de Meent}, Maarten and {Muratore}, Martina and {Nardini}, Germano and {Pani}, Paolo and {Peloso}, Marco and {Pieroni}, Mauro and {Pound}, Adam and {Quelquejay-Leclere}, Hippolyte and {Ricciardone}, Angelo and {Rossi}, Elena Maria and {Sartirana}, Andrea and {Savalle}, Etienne and {Sberna}, Laura and {Sesana}, Alberto and {Shoemaker}, Deirdre and {Slutsky}, Jacob and {Sotiriou}, Thomas and {Speri}, Lorenzo and {Staab}, Martin and {Steer}, Dani{\`e}le and {Tamanini}, Nicola and {Tasinato}, Gianmassimo and {Torrado}, Jesus and {Torres-Orjuela}, Alejandro and {Toubiana}, Alexandre and {Vallisneri}, Michele and {Vecchio}, Alberto and {Volonteri}, Marta and {Yagi}, Kent and {Zwick}, Lorenz},
        title = "{LISA Definition Study Report}",
      journal = {arXiv e-prints},
         year = 2024,
        month = feb,
          eid = {arXiv:2402.07571},
        pages = {arXiv:2402.07571},
          doi = {10.48550/arXiv.2402.07571},
archivePrefix = {arXiv},
       eprint = {2402.07571},
 primaryClass = {astro-ph.CO},
       adsurl = {https://ui.adsabs.harvard.edu/abs/2024arXiv240207571C}
}

@ARTICLE{Talts+2018,
       author = {{Talts}, Sean and {Betancourt}, Michael and {Simpson}, Daniel and {Vehtari}, Aki and {Gelman}, Andrew},
        title = "{Validating Bayesian Inference Algorithms with Simulation-Based Calibration}",
      journal = {arXiv e-prints},
         year = 2018,
        month = apr,
          eid = {arXiv:1804.06788},
        pages = {arXiv:1804.06788},
          doi = {10.48550/arXiv.1804.06788},
archivePrefix = {arXiv},
       eprint = {1804.06788},
 primaryClass = {stat.ME},
       adsurl = {https://ui.adsabs.harvard.edu/abs/2018arXiv180406788T}
}

@ARTICLE{SNLE_2018,
       author = {{Papamakarios}, George and {Sterratt}, David C. and {Murray}, Iain},
        title = "{Sequential Neural Likelihood: Fast Likelihood-free Inference with Autoregressive Flows}",
      journal = {arXiv e-prints},
         year = 2018,
        month = may,
          eid = {arXiv:1805.07226},
        pages = {arXiv:1805.07226},
          doi = {10.48550/arXiv.1805.07226},
archivePrefix = {arXiv},
       eprint = {1805.07226},
 primaryClass = {stat.ML},
       adsurl = {https://ui.adsabs.harvard.edu/abs/2018arXiv180507226P}
}

@ARTICLE{MWdisclength_2008,
       author = {{Juric}, Mario and {Ivezic}, Zeljko and {Brooks}, Alyson and {Lupton}, Robert H. and {Schlegel}, David and {Finkbeiner}, Douglas and {Padmanabhan}, Nikhil and {Bond}, Nicholas and {Sesar}, Branimir and {Rockosi}, Constance M. and {Knapp}, Gillian R. and {Gunn}, James E. and {Sumi}, Takahiro and {Schneider}, Donald P. and {Barentine}, J.~C. and {Brewington}, Howard J. and {Brinkmann}, J. and {Fukugita}, Masataka and {Harvanek}, Michael and {Kleinman}, S.~J. and {Krzesinski}, Jurek and {Long}, Dan and {Neilsen}, Jr., Eric H. and {Nitta}, Atsuko and {Snedden}, Stephanie A. and {York}, Donald G.},
        title = "{The Milky Way Tomography with SDSS. I. Stellar Number Density Distribution}",
      journal = {\apj},
         year = 2008,
        month = feb,
       volume = {673},
       number = {2},
        pages = {864-914},
          doi = {10.1086/523619},
archivePrefix = {arXiv},
       eprint = {astro-ph/0510520},
 primaryClass = {astro-ph},
       adsurl = {https://ui.adsabs.harvard.edu/abs/2008ApJ...673..864J}
}

@ARTICLE{IMNN_2018,
       author = {{Charnock}, Tom and {Lavaux}, Guilhem and {Wandelt}, Benjamin D.},
        title = "{Automatic physical inference with information maximizing neural networks}",
      journal = {\prd},
         year = 2018,
        month = apr,
       volume = {97},
       number = {8},
          eid = {083004},
        pages = {083004},
          doi = {10.1103/PhysRevD.97.083004},
archivePrefix = {arXiv},
       eprint = {1802.03537},
 primaryClass = {astro-ph.IM},
       adsurl = {https://ui.adsabs.harvard.edu/abs/2018PhRvD..97h3004C}
}

@article{AnauMontel:2024flo,
    author = "Anau Montel, Noemi and Alvey, James and Weniger, Christoph",
    title = "{Tests for model misspecification in simulation-based inference: From local distortions to global model checks}",
    eprint = "2412.15100",
    archivePrefix = "arXiv",
    primaryClass = "astro-ph.IM",
    doi = "10.1103/PhysRevD.111.083013",
    journal = "Phys. Rev. D",
    volume = "111",
    number = "8",
    pages = "083013",
    year = "2025"
}

@article{Cole:2025sqo,
    author = "Cole, Philippa S. and Alvey, James and Speri, Lorenzo and Weniger, Christoph and Bhardwaj, Uddipta and Gerosa, Davide and Bertone, Gianfranco",
    title = "{Sequential simulation-based inference for extreme mass ratio inspirals}",
    eprint = "2505.16795",
    archivePrefix = "arXiv",
    primaryClass = "gr-qc",
    month = "5",
    year = "2025"
}

@article{Alvey:2023npw,
    author = "Alvey, James and Bhardwaj, Uddipta and Domcke, Valerie and Pieroni, Mauro and Weniger, Christoph",
    title = "{Simulation-based inference for stochastic gravitational wave background data analysis}",
    eprint = "2309.07954",
    archivePrefix = "arXiv",
    primaryClass = "gr-qc",
    reportNumber = "CERN-TH-2023-167",
    doi = "10.1103/PhysRevD.109.083008",
    journal = "Phys. Rev. D",
    volume = "109",
    number = "8",
    pages = "083008",
    year = "2024"
}

@ARTICLE{Perigrine_2023,
       author = {{Bhardwaj}, Uddipta and {Alvey}, James and {Miller}, Benjamin Kurt and {Nissanke}, Samaya and {Weniger}, Christoph},
        title = "{Sequential simulation-based inference for gravitational wave signals}",
      journal = {\prd},
         year = 2023,
        month = aug,
       volume = {108},
       number = {4},
          eid = {042004},
        pages = {042004},
          doi = {10.1103/PhysRevD.108.042004},
archivePrefix = {arXiv},
       eprint = {2304.02035},
 primaryClass = {gr-qc},
       adsurl = {https://ui.adsabs.harvard.edu/abs/2023PhRvD.108d2004B}
}

@ARTICLE{RobustSummaries_ModelMispec_2023,
       author = {{Huang}, Daolang and {Bharti}, Ayush and {Souza}, Amauri and {Acerbi}, Luigi and {Kaski}, Samuel},
        title = "{Learning Robust Statistics for Simulation-based Inference under Model Misspecification}",
      journal = {arXiv e-prints},
         year = 2023,
        month = may,
          eid = {arXiv:2305.15871},
        pages = {arXiv:2305.15871},
          doi = {10.48550/arXiv.2305.15871},
archivePrefix = {arXiv},
       eprint = {2305.15871},
 primaryClass = {stat.ML},
       adsurl = {https://ui.adsabs.harvard.edu/abs/2023arXiv230515871H}
}

@ARTICLE{Model_mispec_SBI_2021,
       author = {{Schmitt}, Marvin and {B{\"u}rkner}, Paul-Christian and {K{\"o}the}, Ullrich and {Radev}, Stefan T.},
        title = "{Detecting Model Misspecification in Amortized Bayesian Inference with Neural Networks}",
      journal = {arXiv e-prints},
         year = 2021,
        month = dec,
          eid = {arXiv:2112.08866},
        pages = {arXiv:2112.08866},
          doi = {10.48550/arXiv.2112.08866},
archivePrefix = {arXiv},
       eprint = {2112.08866},
 primaryClass = {stat.ME},
       adsurl = {https://ui.adsabs.harvard.edu/abs/2021arXiv211208866S}
}

@ARTICLE{Model_misspec_2023,
       author = {{Huang}, Daolang and {Bharti}, Ayush and {Souza}, Amauri and {Acerbi}, Luigi and {Kaski}, Samuel},
        title = "{Learning Robust Statistics for Simulation-based Inference under Model Misspecification}",
      journal = {arXiv e-prints},
         year = 2023,
        month = may,
          eid = {arXiv:2305.15871},
        pages = {arXiv:2305.15871},
          doi = {10.48550/arXiv.2305.15871},
archivePrefix = {arXiv},
       eprint = {2305.15871},
 primaryClass = {stat.ML},
       adsurl = {https://ui.adsabs.harvard.edu/abs/2023arXiv230515871H}
}

@ARTICLE{Model_misspec_2024,
       author = {{Wehenkel}, Antoine and {Gamella}, Juan L. and {Sener}, Ozan and {Behrmann}, Jens and {Sapiro}, Guillermo and {Jacobsen}, J{\"o}rn-Henrik and {Cuturi}, Marco},
        title = "{Addressing Misspecification in Simulation-based Inference through Data-driven Calibration}",
      journal = {arXiv e-prints},
         year = 2024,
        month = may,
          eid = {arXiv:2405.08719},
        pages = {arXiv:2405.08719},
          doi = {10.48550/arXiv.2405.08719},
archivePrefix = {arXiv},
       eprint = {2405.08719},
 primaryClass = {stat.ML},
       adsurl = {https://ui.adsabs.harvard.edu/abs/2024arXiv240508719W}
}

@ARTICLE{LISAInstruments_2023,
       author = {{Bayle}, Jean-Baptiste and {Hartwig}, Olaf},
        title = "{Unified model for the LISA measurements and instrument simulations}",
      journal = {\prd},
         year = 2023,
        month = apr,
       volume = {107},
       number = {8},
          eid = {083019},
        pages = {083019},
          doi = {10.1103/PhysRevD.107.083019},
archivePrefix = {arXiv},
       eprint = {2212.05351},
 primaryClass = {gr-qc},
       adsurl = {https://ui.adsabs.harvard.edu/abs/2023PhRvD.107h3019B}
}

@ARTICLE{PCA_pop_TaylorGerosa2018,
       author = {{Taylor}, Stephen R. and {Gerosa}, Davide},
        title = "{Mining gravitational-wave catalogs to understand binary stellar evolution: A new hierarchical Bayesian framework}",
      journal = {\prd},
         year = 2018,
        month = oct,
       volume = {98},
       number = {8},
          eid = {083017},
        pages = {083017},
          doi = {10.1103/PhysRevD.98.083017},
archivePrefix = {arXiv},
       eprint = {1806.08365},
 primaryClass = {astro-ph.HE},
       adsurl = {https://ui.adsabs.harvard.edu/abs/2018PhRvD..98h3017T}
}

@misc{natalia_code_2022,
  author       = {Natalia Korsakova},
  title        = {generate\_galaxy: Galaxy population synthesis code},
  year         = {2022},
  url          = {https://github.com/NataliaKor/generate_galaxy},
  note         = {Accessed: 2025-06-05}
}

@article{GBGPU-ref4,
  title={Detecting hierarchical stellar systems with LISA},
  author={Robson, Travis and Cornish, Neil J and Tamanini, Nicola and Toonen, Silvia},
  journal={Physical Review D},
  volume={98},
  number={6},
  pages={064012},
  year={2018},
  publisher={APS}
}

@misc{Korol+2022_code,
  author       = {Korol, Valeriya and Hallakoun, Na’ama and Toonen, Silvia and Karnesis, Nikolao},
  title        = {{Observationally-driven population of Galactic binaries for LISA}},
  year         = {2022},
  howpublished = {\url{https://gitlab.in2p3.fr/korol/observationally-driven-population-of-galactic-binaries}},
}

@misc{LDC2022,
  author       = {{LISA Data Challenge working group}},
  title        = {{LISA Data Challenge software}},
  year         = {2022},
  publisher    = {Zenodo},
  doi          = {10.5281/zenodo.7332221},
  url          = {https://doi.org/10.5281/zenodo.7332221},
  note         = {Accessed via Zenodo}
}

@software{katz2022gbgpu,
  author = {Katz, Michael L.},
  title = {mikekatz04/GBGPU: First official public release!},
  version = {1.0.0},
  publisher = {Zenodo},
  doi = {10.5281/zenodo.6500434},
  year = {2022},
  url = {https://doi.org/10.5281/zenodo.6500434}
}

@ARTICLE{Katz+2022,
       author = {{Katz}, Michael L. and {Danielski}, Camilla and {Karnesis}, Nikolaos and {Korol}, Valeriya and {Tamanini}, Nicola and {Cornish}, Neil J. and {Littenberg}, Tyson B.},
        title = "{Bayesian characterization of circumbinary sub-stellar objects with LISA}",
      journal = {\mnras},
         year = 2022,
        month = nov,
       volume = {517},
       number = {1},
        pages = {697-711},
          doi = {10.1093/mnras/stac2555},
archivePrefix = {arXiv},
       eprint = {2205.03461},
 primaryClass = {astro-ph.EP},
       adsurl = {https://ui.adsabs.harvard.edu/abs/2022MNRAS.517..697K}
}

@ARTICLE{Cornish_Littenberg_2017,
       author = {{Cornish}, Neil J. and {Littenberg}, Tyson B.},
        title = "{Tests of Bayesian model selection techniques for gravitational wave astronomy}",
      journal = {\prd},
         year = 2007,
        month = oct,
       volume = {76},
       number = {8},
          eid = {083006},
        pages = {083006},
          doi = {10.1103/PhysRevD.76.083006},
archivePrefix = {arXiv},
       eprint = {0704.1808},
 primaryClass = {gr-qc},
       adsurl = {https://ui.adsabs.harvard.edu/abs/2007PhRvD..76h3006C}
}

@article{Gravity:2019,
    author = "Abuter, R. and others",
    collaboration = "Gravity",
    title = "{A geometric distance measurement to the Galactic center black hole with 0.3\% uncertainty}",
    eprint = "1904.05721",
    archivePrefix = "arXiv",
    primaryClass = "astro-ph.GA",
    doi = "10.1051/0004-6361/201935656",
    journal = "Astron. Astrophys.",
    volume = "625",
    year = "2019"
}

@ARTICLE{Gaia2018,
       author = {{Gaia Collaboration} and {Brown}, A.~G.~A. and {Vallenari}, A. and others},
        title = "{Gaia Data Release 2. Summary of the contents and survey properties}",
      journal = {\aap},
         year = 2018,
        month = aug,
       volume = {616},
          eid = {A1},
        pages = {A1},
          doi = {10.1051/0004-6361/201833051},
archivePrefix = {arXiv},
       eprint = {1804.09365},
 primaryClass = {astro-ph.GA},
       adsurl = {https://ui.adsabs.harvard.edu/abs/2018A&A...616A...1G}
}

@article{SBIfrontier:Kranmer+2020,
    author = {Kyle Cranmer  and Johann Brehmer  and Gilles Louppe },
    title = {The frontier of simulation-based inference},
    journal = {Proceedings of the National Academy of Sciences},
    volume = {117},
    number = {48},
    pages = {30055-30062},
    year = {2020},
    doi = {10.1073/pnas.1912789117},
    URL = {https://www.pnas.org/doi/abs/10.1073/pnas.1912789117},
    eprint = {https://www.pnas.org/doi/pdf/10.1073/pnas.1912789117}
}

@article{LISA:2017pwj,
    author = "Amaro-Seoane, Pau and others",
    collaboration = "LISA",
    title = "{Laser Interferometer Space Antenna}",
    eprint = "1702.00786",
    archivePrefix = "arXiv",
    primaryClass = "astro-ph.IM",
    month = "2",
    year = "2017"
}

@article{Houba:2024ysm,
    author = "Houba, Niklas and Bayle, Jean-Baptiste and Vallisneri, Michele",
    title = "{Robust Bayesian inference with gapped LISA data using all-in-one TDI-$\infty$}",
    eprint = "2412.20793",
    archivePrefix = "arXiv",
    primaryClass = "astro-ph.IM",
    month = "12",
    year = "2024"
}

@article{Baghi:2019eqo,
    author = "Baghi, Quentin and Thorpe, Ira and Slutsky, Jacob and Baker, John and Dal Canton, Tito and Korsakova, Natalia and Karnesis, Nikos",
    title = "{Gravitational-wave parameter estimation with gaps in LISA: a Bayesian data augmentation method}",
    eprint = "1907.04747",
    archivePrefix = "arXiv",
    primaryClass = "gr-qc",
    doi = "10.1103/PhysRevD.100.022003",
    journal = "Phys. Rev. D",
    volume = "100",
    number = "2",
    pages = "022003",
    year = "2019"
}

@article{Sesana:2016ljz,
    author = "Sesana, Alberto",
    title = "{Prospects for Multiband Gravitational-Wave Astronomy after GW150914}",
    eprint = "1602.06951",
    archivePrefix = "arXiv",
    primaryClass = "gr-qc",
    doi = "10.1103/PhysRevLett.116.231102",
    journal = "Phys. Rev. Lett.",
    volume = "116",
    number = "23",
    pages = "231102",
    year = "2016"
}

@article{Barausse:2023yrx,
    author = "Barausse, Enrico and Dey, Kallol and Crisostomi, Marco and Panayada, Akshay and Marsat, Sylvain and Basak, Soumen",
    title = "{Implications of the pulsar timing array detections for massive black hole mergers in the LISA band}",
    eprint = "2307.12245",
    archivePrefix = "arXiv",
    primaryClass = "astro-ph.GA",
    doi = "10.1103/PhysRevD.108.103034",
    journal = "Phys. Rev. D",
    volume = "108",
    number = "10",
    pages = "103034",
    year = "2023"
}

@article{Crisostomi:2023tle,
    author = "Crisostomi, Marco and Dey, Kallol and Barausse, Enrico and Trotta, Roberto",
    title = "{Neural posterior estimation with guaranteed exact coverage: The ringdown of GW150914}",
    eprint = "2305.18528",
    archivePrefix = "arXiv",
    primaryClass = "gr-qc",
    doi = "10.1103/PhysRevD.108.044029",
    journal = "Phys. Rev. D",
    volume = "108",
    number = "4",
    pages = "044029",
    year = "2023"
}

@article{Dey:2021dem,
    author = "Dey, Kallol and Karnesis, Nikolaos and Toubiana, Alexandre and Barausse, Enrico and Korsakova, Natalia and Baghi, Quentin and Basak, Soumen",
    title = "{Effect of data gaps on the detectability and parameter estimation of massive black hole binaries with LISA}",
    eprint = "2104.12646",
    archivePrefix = "arXiv",
    primaryClass = "gr-qc",
    doi = "10.1103/PhysRevD.104.044035",
    journal = "Phys. Rev. D",
    volume = "104",
    number = "4",
    pages = "044035",
    year = "2021"
}

@article{Babak:2017tow,
    author = "Babak, Stanislav and Gair, Jonathan and Sesana, Alberto and Barausse, Enrico and Sopuerta, Carlos F. and Berry, Christopher P. L. and Berti, Emanuele and Amaro-Seoane, Pau and Petiteau, Antoine and Klein, Antoine",
    title = "{Science with the space-based interferometer LISA. V: Extreme mass-ratio inspirals}",
    eprint = "1703.09722",
    archivePrefix = "arXiv",
    primaryClass = "gr-qc",
    doi = "10.1103/PhysRevD.95.103012",
    journal = "Phys. Rev. D",
    volume = "95",
    number = "10",
    pages = "103012",
    year = "2017"
}

@article{Klein:2015hvg,
    author = "Klein, Antoine and others",
    title = "{Science with the space-based interferometer eLISA: Supermassive black hole binaries}",
    eprint = "1511.05581",
    archivePrefix = "arXiv",
    primaryClass = "gr-qc",
    doi = "10.1103/PhysRevD.93.024003",
    journal = "Phys. Rev. D",
    volume = "93",
    number = "2",
    pages = "024003",
    year = "2016"
}

@article{LISA:2024hlh,
    author = "Colpi, Monica and others",
    collaboration = "LISA",
    title = "{LISA Definition Study Report}",
    eprint = "2402.07571",
    archivePrefix = "arXiv",
    primaryClass = "astro-ph.CO",
    month = "2",
    year = "2024"
}

@article{MAF,
  title={Masked autoregressive flow for density estimation},
  author={Papamakarios, George and Pavlakou, Theo and Murray, Iain},
  journal={Advances in neural information processing systems},
  volume={30},
  year={2017}
}

@ARTICLE{Srinivasan+2023,
       author = {{Srinivasan}, Rahul and {Lamberts}, Astrid and {Bizouard}, Marie Anne and {Bruel}, Tristan and {Mastrogiovanni}, Simone},
        title = "{Understanding the progenitor formation galaxies of merging binary black holes}",
      journal = {\mnras},
         year = 2023,
        month = sep,
       volume = {524},
       number = {1},
        pages = {60-75},
          doi = {10.1093/mnras/stad1825},
archivePrefix = {arXiv},
       eprint = {2303.04017},
 primaryClass = {astro-ph.HE},
       adsurl = {https://ui.adsabs.harvard.edu/abs/2023MNRAS.524...60S}
}

@ARTICLE{Iorio+2023,
       author = {{Iorio}, Giuliano and {Mapelli}, Michela and {Costa}, Guglielmo and {Spera}, Mario and {Escobar}, Gast{\'o}n J. and {Sgalletta}, Cecilia and {Trani}, Alessandro A. and {Korb}, Erika and {Santoliquido}, Filippo and {Dall'Amico}, Marco and {Gaspari}, Nicola and {Bressan}, Alessandro},
        title = "{Compact object mergers: exploring uncertainties from stellar and binary evolution with SEVN}",
      journal = {\mnras},
         year = 2023,
        month = sep,
       volume = {524},
       number = {1},
        pages = {426-470},
          doi = {10.1093/mnras/stad1630},
archivePrefix = {arXiv},
       eprint = {2211.11774},
 primaryClass = {astro-ph.HE},
       adsurl = {https://ui.adsabs.harvard.edu/abs/2023MNRAS.524..426I}
}

@ARTICLE{Breivik+2020,
       author = {{Breivik}, Katelyn and {Mingarelli}, Chiara M.~F. and {Larson}, Shane L.},
        title = "{Constraining Galactic Structure with the LISA White Dwarf Foreground}",
      journal = {\apj},
         year = 2020,
        month = sep,
       volume = {901},
       number = {1},
          eid = {4},
        pages = {4},
          doi = {10.3847/1538-4357/abab99},
archivePrefix = {arXiv},
       eprint = {1912.02200},
 primaryClass = {astro-ph.GA},
       adsurl = {https://ui.adsabs.harvard.edu/abs/2020ApJ...901....4B}
}

@ARTICLE{Leyde+2024,
       author = {{Leyde}, Konstantin and {Green}, Stephen R. and {Toubiana}, Alexandre and {Gair}, Jonathan},
        title = "{Gravitational wave populations and cosmology with neural posterior estimation}",
      journal = {\prd},
         year = 2024,
        month = mar,
       volume = {109},
       number = {6},
          eid = {064056},
        pages = {064056},
          doi = {10.1103/PhysRevD.109.064056},
archivePrefix = {arXiv},
       eprint = {2311.12093},
 primaryClass = {gr-qc},
       adsurl = {https://ui.adsabs.harvard.edu/abs/2024PhRvD.109f4056L}
}

@ARTICLE{Mold+2025,
       author = {{Mould}, Matthew and {Wolfe}, Noah E. and {Vitale}, Salvatore},
        title = "{Rapid inference and comparison of gravitational-wave population models with neural variational posteriors}",
      journal = {arXiv e-prints},
         year = 2025,
        month = apr,
          eid = {arXiv:2504.07197},
        pages = {arXiv:2504.07197},
          doi = {10.48550/arXiv.2504.07197},
archivePrefix = {arXiv},
       eprint = {2504.07197},
 primaryClass = {astro-ph.IM},
       adsurl = {https://ui.adsabs.harvard.edu/abs/2025arXiv250407197M}
}

@ARTICLE{Adams+2012,
       author = {{Adams}, Matthew R. and {Cornish}, Neil J. and {Littenberg}, Tyson B.},
        title = "{Astrophysical model selection in gravitational wave astronomy}",
      journal = {\prd},
         year = 2012,
        month = dec,
       volume = {86},
       number = {12},
          eid = {124032},
        pages = {124032},
          doi = {10.1103/PhysRevD.86.124032},
archivePrefix = {arXiv},
       eprint = {1209.6286},
 primaryClass = {gr-qc},
       adsurl = {https://ui.adsabs.harvard.edu/abs/2012PhRvD..86l4032A}
}

\end{document}